\documentclass{article}

\usepackage[english]{babel}
\usepackage{natbib}
\usepackage[letterpaper,top=2cm,bottom=2cm,left=3cm,right=3cm,marginparwidth=1.75cm]{geometry}
\usepackage{amsmath, amsthm, amssymb}
\usepackage{bbm}
\usepackage{graphicx}
\usepackage[colorlinks=true, allcolors=blue]{hyperref}
\usepackage{subcaption}
\usepackage{mathtools}
\usepackage{wrapfig}
\usepackage{array}
\usepackage{dsfont}
\usepackage[linesnumbered,ruled]{algorithm2e}
\usepackage{algorithmic}

\DeclareMathOperator*{\argmin}{argmin}
\DeclareMathOperator*{\argmax}{argmax}

\title{A Spatio-Temporal Machine Learning Model for Mortgage Credit Risk: Default Probabilities and Loan Portfolios}
\author{
  Pascal Kündig\footnotemark[1]  \footnotemark[3] \footnotemark[4]\\
  \and
  Fabio Sigrist\footnotemark[2] \footnotemark[1]  
}

\begin{document}
\date{}
\maketitle

\begin{abstract}
We introduce a novel machine learning model for credit risk by combining tree-boosting with a latent spatio-temporal Gaussian process model accounting for frailty correlation. This allows for modeling non-linearities and interactions among predictor variables in a flexible data-driven manner and for accounting for spatio-temporal variation that is not explained by observable predictor variables. We also show how estimation and prediction can be done in a computationally efficient manner. In an application to a large U.S. mortgage credit risk data set, we find that both predictive default probabilities for individual loans and predictive loan portfolio loss distributions obtained with our novel approach are more accurate compared to conventional independent linear hazard models and also linear spatio-temporal models. Using interpretability tools for machine learning models, we find that the likely reasons for this outperformance are strong interaction and non-linear effects in the predictor variables and the presence of spatio-temporal frailty effects. 
\end{abstract}


\footnotetext[1]{Lucerne University of Applied Sciences and Arts}
\footnotetext[2]{Seminar for Statistics, ETH Zurich}
\footnotetext[3]{University of Basel}
\footnotetext[4]{Corresponding author: pascal.kuendig@gmail.com}

\section{Introduction}
Accurately modeling credit default risk is a challenging and critical task for financial institutions, investors, regulators, and policy makers. Traditionally, linear models such as linear discriminant analysis, logistic regression, and linear discrete hazard models have been widely used to model bankruptcy probabilities in corporate credit risk research \citep{altman1968financial, zmijewski1984methodological, shumway2001forecasting}. These linear models are also frequently applied in the retail credit risk literature \citep{crook2007recent, thomas2017credit} for predicting default probabilities of various loan types, including mortgage loans \citep{agarwal2012thy}, credit card loans \citep{bellotti2009support, bellotti2009credit, bellotti2013forecasting}, and personal loans \citep{banasik1999not, malik2010modelling}.


For modeling corporate credit risk, linear hazard models have been extended to account for temporal frailty correlation that cannot be captured by observable predictor variables \citep{duffie2009frailty, koopman2011modeling}. Similarly, for loans associated with a specific spatial location, such as mortgages and loans to small and medium-sized enterprises (SMEs), spatial dependence likely exists among loans that cannot be fully explained by observable predictor variables. Specifically, for mortgage credit risk, there are the following potential reasons for the presence of frailty correlation. First, local weather extremes such as wildfires \citep{issler2024housing} and hurricanes \citep{kousky2020flood, calabrese2024impacts} have an impact on mortgage credit risk that leads to correlations among defaults. Furthermore, local economic conditions that often cannot be measured using observable predictor variables likely impact default risks. For instance, regions heavily reliant on specific industries are vulnerable to concentrated job losses and income reductions during industry downturns. This can lead to widespread local defaults as borrowers face diminished ability to make payments. In addition, house prices are often locally interconnected, and when local house prices fall, many borrowers in an area can enter negative equity. This can lead to foreclosure contagion which further lowers property values. \citet{harding2009contagion} show that neighborhood foreclosures have the greatest contagion effect on house prices within a 300-foot radius. This highlights the importance of accounting for local dependencies between properties in immediate neighborhoods. 

To account for spatial frailty correlation, several linear models have been proposed in the SME credit risk literature \citep{fernandes2016spatial, agosto2019spatial, calabrese2019birds, medina2022spatial}. 
Recently, \citet{berloco2023forecasting} introduced a model accounting for spatio-temporal correlation among firms while retaining a linear functional form for the predictor variables. Early approaches to account for spatial dependence in the mortgage credit risk literature include the use of independent and identically distributed frailty variables for metropolitan statistical areas (MSAs) \citep{kau2011analysis} and ZIP code concentration variables \citep{agarwal2012thy}. \citet{zhu2014modeling} demonstrate that spatial probit models offer superior prediction accuracy for mortgage defaults compared to independent probit models. \citet{calabrese2020spatial} extended this spatial methodology to linear hazard models. Further research on spatial frailty correlation for mortgage credit risk include \citet{babii2019commercial, calabrese2024impacts}, and \citet{medina2023spacetime} who introduce a linear spatio-temporal model to predict mortgage prepayment. Additional prior works on spatial frailty modeling outside the credit risk literature include \citet{li2002modeling}, \citet{banerjee2003semiparametric}, \citet{hanson2012bayesian}, and \citet{vcivzek2017neighbours}.

Prior research has also investigated the effectiveness of machine learning methods like random forests, tree-boosting, and deep neural networks in modeling corporate credit risk \citep{barboza2017machine, zikeba2016, sigrist2019grabit, sigrist2023machine, cheraghali2025sme} and retail credit risk \citep{lessmann2015benchmarking, fitzpatrick2016empirical, xia2017boosted, thomas2017credit}. These non-linear machine learning methods often outperform traditional linear models in prediction accuracy by capturing complex interactions and non-linear effects in the predictor variables. For instance, \citet{fitzpatrick2016empirical} find that tree-boosting is preferred over logistic regression for predicting mortgage defaults.
To the best of our knowledge, there exists no prior work that uses state-of-the-art machine learning models and explicitly accounts for spatio-temporal frailty correlation for modeling credit default risk in general and mortgage risk in particular.

In this article, we introduce a novel credit risk modeling approach which combines tree-boosting with a latent spatio-temporal Gaussian process. This allows for modeling non-linear and interaction effects of predictor variables as well as for accounting for spatio-temporal frailty correlation among loans which is not accounted for by the observable predictor variables. It is likely that not all relationships are linear, and more realistic non-linear models allow for gaining a better understanding of default mechanisms and for generating more accurate predictions. Furthermore, a space-time Gaussian process integrated in the model allows for accounting for frailty correlation and for generating spatially and temporally varying frailty default risk maps. Besides providing valuable insights on its own, this can improve the prediction accuracy of default probabilities and loan portfolio distributions. To the best of our knowledge, this is the first work combining space-time Gaussian processes with tree-boosting, also outside of the credit risk literature.

We apply our proposed model to a large U.S. mortgage data set and compare it to a linear hazard, a linear spatial, and a linear spatio-temporal model. We first observe that incorporating spatial and spatio-temporal frailty correlation in a linear model improves the prediction accuracy of default probabilities compared to an independent linear hazard model. Our newly proposed methodology, which extends spatial and spatio-temporal linear models using tree-boosting, leads to further improvements in prediction accuracy of default probabilities. Although the differences in prediction accuracy for individual default probabilities are relatively small, they are nonetheless important, as even small differences are relevant for large portfolios of mortgages. For example, when considering loan portfolios consisting of all active mortgages in our data set, we find that our proposed spatial and spatio-temporal frailty machine learning models yield considerably more accurate predictive 99\% upper quantiles in terms of the quantile loss compared to all linear models, i.e., a linear hazard model without frailty correlation, a linear spatial model, and a linear spatio-temporal frailty model. In particular, the proposed tree-boosted frailty models predict more realistic losses at the beginning and end of the global financial crisis around the year 2008 with, e.g., mean losses of frailty machine learning models being almost US \$1 billion more accurate compared to linear models. Using interpretability tools for machine learning models, we find that there are strong interactions and non-linear effects in the predictor variables which cannot be captured with a linear functional form. In addition, we find that there are sizable spatio-temporal frailty effects.

The remainder of this article is organized as follows. In Section \ref{section:models}, we introduce the methodology, and in Section \ref{section:application}, we apply and compare our methods on a large U.S. mortgage data set. Section \ref{section:conclusion} concludes.

\section{Spatio-temporal frailty correlation and tree-boosting} \label{section:models}

\subsection{Notation and default probabilities}
Our goal is to model default events of $N\in\mathbb{N}^+$ loans. For every loan $i$, $i=1,\dots,N$, we observe predictor variables $X_{ki}\in \mathbb{R}^p$ at discrete times $t_{ki}$, $k=0,\dots,n_i-1$, and a default time $\tau_i\in\{t_{1i},\dots,t_{n_ii},\infty\}$, where $0\leq t_{0i}\leq t_{ki} \leq t_{n_i i}\leq T$, $k=0,\dots,n_i$. I.e., $X_{ki}$ are the predictor variables of loan $i$ observed at time $t_{ki}$, and $\tau_i=t_{ki}>t_{0i}$ means that loan $i$ has defaulted in the interval $(t_{k-1i},t_{ki}]$. Further, $t_{0i}$ denotes the time when a loan $i$ enters the set of active loans, $t_{n_i i}$ denotes the last observation time for loan $i$, and $n_i$ is the total number of temporal observations for loan $i$. The last observation time $t_{n_i i}$ can be either the default time $\tau_i$, the time of some other form of exit such as reaching the loan maturity date, or the end of the observation period $T$. In addition, we assume that every loan $i$ has associated spatial coordinates $s_i \in \mathcal{D}\subset \mathbb{R}^2$. 

Let $P\left(\tau_i=t_{k+1i}\middle|\tau_i>t_{ki}\right)$ denote the probability that a loan $i$ defaults in the interval $(t_{ki},t_{k+1i}]$ given that it has not defaulted until time $t_{ki}$. In a traditional independent linear hazard model, it is assumed that 
\begin{equation}\label{prob_lin_mod}
    P\left(\tau_i=t_{k+1i}\middle|\tau_i>t_{ki}\right) = \left(1+e^{-X_{ki}^T \beta}\right)^{-1}, ~~ \beta\in \mathbb{R}^p.
\end{equation}
Assuming independence across space and time conditional on $X_{ki}$, the corresponding likelihood is given by
\begin{equation}\label{lla}
\prod_{i=1}^N\prod_{k=0}^{n_i-1}\left(\mathbbm{1}_{\{\tau_i=t_{k+1i}\}}\left(1+e^{-X_{ki}^T \beta}\right)^{-1} + \mathbbm{1}_{\{\tau_i>t_{k+1i}\}}\left(1-\left(1+e^{-X_{ki}^T \beta}\right)^{-1}\right)\right).
\end{equation}


\subsection{Accounting for spatial and spatio-temporal frailty correlation}
In the following, we first lift the assumption of independence between loan-time observations by introducing latent frailty variables that model spatial and spatio-temporal correlation. Instead of \eqref{prob_lin_mod}, we assume 
\begin{equation}\label{cond_prob}
P\left(\tau_i=t_{k+1i}\middle|\tau_i>t_{ki},b(t_{ki},s_i)\right)=f(F(X_{ki}) + b(t_{ki},s_i)),
\end{equation}
where $f(\cdot)$ is a link function such as $f(x)=\left(1+e^{-x}\right)^{-1}$, $F(\cdot)$ is a function $F:\mathbb{R}^p \mapsto \mathbb{R}$, and the latent variable $b(\cdot,\cdot)$ is a zero-mean Gaussian process \citep{williams2006gaussian} that accounts for spatial or spatio-temporal frailty correlation. Both $b(\cdot,\cdot)$ and $F(\cdot)$ are specified in the following.  

For the latent frailty process $b(t,s)$, we consider two cases: a spatial model and a spatio-temporal model. In the spatial model, the Gaussian process $b(t,s)$ varies over space only and is constant over time. It is defined by a spatial covariance function $\text{Cov}(b(t,s),b(t',s')) = c_{\theta}(s,s')$, $s,s' \in \mathcal{D}$, which depends on a set of parameters $\theta\in\Theta \subset \mathbb{R}^q$. For more information on covariance functions in general, see \citet[Chapter 4]{williams2006gaussian}, and for spatial covariance functions in particular, see \citet[Section 2.3.1]{cressie2015statistics}. In the application below, we use a Matérn covariance function:
\begin{equation}\label{matern_spatial}
    c_{\theta}(s,s')=\sigma_1^2 \frac{2^{\nu -1}}{\Gamma(\nu)}\left(\frac{\sqrt{2\nu}||s-s'||}{\rho_s}\right)^\nu K_\nu\left(\frac{\sqrt{2\nu}||s-s'||}{\rho_s}\right),
\end{equation}
where $\sigma_1^2$ is a marginal variance parameter, $\rho_s$ is a spatial range, or scale, parameter, $\nu$ is a smoothness parameter, $\Gamma(\cdot)$ is the Gamma function, and $K_\nu(\cdot)$ is the modified Bessel function of the second kind. For spatio-temporal models, the Gaussian process varies over both space and time, and it is defined by a space-time covariance function $\text{Cov}(b(t,s),b(t',s')) = c_{\theta}((t,s),(t',s'))$, $(t,s),(t',s') \in [0,T]\times \mathcal{D} \subset \mathbb{R}^3$. For more information on space-time covariance functions, we refer to \citet[Section 6.1]{cressie2011statistics}. In our application, we use an anisotropic spatio-temporal Matérn covariance function: 
\begin{equation}\label{matern_spatio_temp}
    c_{\theta}((t,s),(t',s'))=\sigma_1^2 \frac{2^{\nu -1}}{\Gamma(\nu)}\left(\sqrt{2\nu}||A^{-1}((t,s)-(t',s'))||\right)^\nu K_\nu\left(\sqrt{2\nu}||A^{-1}((t,s)-(t',s'))||\right),
\end{equation}
where $A = \text{diag}(\rho_t,\rho_s,\rho_s)$ is a diagonal matrix containing a temporal and spatial range parameters $\rho_t$ and $\rho_s$, respectively. We mainly choose this space-time covariance function since it is a parsimonious model that allows for computationally efficient inference when using a Vecchia approximation and a correlation-based neighbor selection approach \citep{kang2023correlation}, see below in Section \ref{model_details} for more details.

The default probability $P\left(\tau_i=t_{k+1i}\middle|\tau_i>t_{ki},b(t_{ki},s_i)\right)$ in \eqref{cond_prob} is conditional on $b(t_{ki},s_i)$, and the joint marginal likelihood of all loans and time points does not simply factorize as in \eqref{lla}. It is given by
\begin{equation}\label{marg_prob}
\mathcal{L}(F,\theta)=\int \left(\prod_{i=1}^N\prod_{k=0}^{n_i-1}\mathcal{L}_{ki}(F(X_{ki}),b(t_{ki},s_i))\right)    p(b|\theta)db,
\end{equation}
where $b$ and $F$ denote the stacked evaluation of $b(\cdot,\cdot)$ and $F(\cdot)$ at all data points, i.e., 
\begin{equation*}
    \begin{split}
        b &= (b(t_{01},s_1),\dots,b(t_{n_1-11},s_1),b(t_{02},s_2),\dots,b(t_{n_2-12},s_2),\dots,b(t_{0N},s_N),\dots,b(t_{n_N-1N},s_N))^T,\\
        F &= (F(X_{01}),\dots,F(X_{n_1-11}),F(X_{02}),\dots,F(X_{n_2-12}),\dots,F(X_{0N}),\dots,F(X_{n_N-1N}))^T,
    \end{split}
\end{equation*}
$p(b|\theta)$ denotes the density of $b$, and $\mathcal{L}_{ki}(F(X_{ki}),b(t_{ki},s_i))$ is defined as
\begin{equation*}
\mathcal{L}_{ki}(F(X_{ki}),b(t_{ki},s_i)) = \mathbbm{1}_{\{\tau_i=t_{k+1i}\}}f(F(X_{ki}) + b(t_{ki},s_i)) + \mathbbm{1}_{\{\tau_i> t_{k+1i}\}}\left(1-f(F(X_{ki}) + b(t_{ki},s_i))\right).
\end{equation*}
If $F(\cdot)$ is a linear function, $F(X_{ki}) =X_{ki}^T\beta $, we refer to a model as defined in \eqref{cond_prob} as a ``linear spatial model" when the latent Gaussian process varies over space only and as a ``linear spatio-temporal model" when the Gaussian process varies over both space and time.

\subsection{A tree-boosted spatio-temporal Gaussian process model}
In the following, we show how to relax the linearity assumption for the fixed effects predictor variable function $F(\cdot)$ using tree-boosting in models with latent Gaussian processes. Tree-boosting \citep{friedman2000additive, friedman2001greedy, buhlmann2007boosting, sigrist2018gradient} is a machine learning technique that often achieves superior prediction accuracy on tabular data sets \citep{nielsen2016tree, shwartz2021tabular, januschowski2022forecasting, grinsztajn2022tree}. For instance, recently \citet{grinsztajn2022tree} showed that tree-boosting outperforms random forest and various state-of-the-art deep neural networks on a large collection of data sets. \citet{sigrist2023machine} and \citet{cheraghali2025sme} find similar results for credit risk data. 

We assume the model specified in \eqref{cond_prob} in the subsequent paragraph. In addition, we assume that $F(\cdot)$ is a function in a normed function space $\mathcal{H}$ that is the linear span of a set $\mathcal{S}$ of base learners $f_j(\cdot):\mathbb{R}^p\rightarrow \mathbb{R}$, which consist of regression trees \citep{breiman1984classification} in this article. Our goal is then to find a joint minimizer for $F(\cdot)\in\mathcal{H}$ and $\theta\in\Theta$ of the functional obtained when plugging $F(\cdot)$ and $\theta$ into the negative log-likelihood:
\begin{equation}\label{optim_def}
(\hat F(\cdot), \hat \theta) =\argmin_{(F(\cdot),\theta) \in (\mathcal{H},\Theta)}-\log\left(\mathcal{L}(F,\theta)\right)\Big|_{F=F(X)},
\end{equation}
where $F(X)$ is the row-wise evaluation of $F(\cdot)$ at $X\in\mathbb{R}^{n\times p}$, which is the matrix containing predictor variables for all observations, $n=\sum_{i=1}^N n_i$, and $\mathcal{L}(F,\theta)$ is given in \eqref{marg_prob}. The minimization of the empirical risk functional in \eqref{optim_def} is done iteratively using the latent Gaussian model boosting (LaGaBoost) algorithm given in Algorithm \ref{LaGaBoost}, which performs a form of functional gradient descent. In detail, this algorithm iterates between, first, finding a maximum for $\theta$ of $\mathcal{L}(F_{m-1},\theta)$ conditional on the current estimate $F_{m-1}(\cdot)$ and, second, updating the ensemble of trees $F(\cdot)$ using one functional gradient descent step given the current estimates $\theta_m$ and $F_{m-1}(\cdot)$. Specifically, the boosting update $f_m(\cdot)$ in iteration $m$ is given by the least squares approximation to the vector obtained when evaluating the negative functional gradient of the functional defined in \eqref{optim_def} at $(F_{m-1}(\cdot), I_{X_{ki}}(\cdot))$, where $I_{X_{ki}}(\cdot)$ are indicator functions which equal $1$ at $X_{ki}$ and $0$ otherwise. Equivalently, $f_m(\cdot)$ is the minimizer of a first-order functional Taylor approximation of the functional in \eqref{optim_def} with $F(\cdot) = F_{m-1}(\cdot) + f(\cdot)$ around $F_{m-1}(\cdot)$ with an $L^2$ penalty on $f(\cdot)$ evaluated at $(X_{ki})$. For more details on the LaGaBoost algorithm, e.g., the calculation of functional gradients; see \citet{sigrist2022gaussian, sigrist2023latent}. 

\begin{algorithm}[ht!]
	\SetKwInOut{Input}{Input}
	\SetKwInOut{Output}{Output}
	\Input{Initial values $\theta_0\in\Theta$, learning rate $\nu>0$, number of boosting iterations $M\in\mathbb{N}$}
	\Output{Function $\hat F(\cdot) = F_{M}(\cdot)$ and parameters $\hat \theta = \theta_M$}
	\caption{LaGaBoost: Latent Gaussian model Boosting}\label{LaGaBoost}
	\begin{algorithmic}[1]
		\STATE Initialize $F_0(\cdot)=\argmax_{c\in\mathbb{R}} \mathcal{L}(c\cdot 1,\theta)$
		\FOR{$m=1$ {\bfseries to} $M$}
		\STATE Find $\theta_m=\underset{\theta\in\Theta}{\argmax} \mathcal{L}(F_{m-1},\theta)$ using a method for convex optimization initialized with $\theta_{m-1}$
		\STATE Find $f_m(\cdot)=\underset{f(\cdot)\in \mathcal{S}}{\argmin}\left\|\frac{\partial\log(\mathcal{L}(F_{m-1},\theta_m))}{\partial F}-f\right\|^2$
		\STATE Update $F_m(\cdot)= F_{m-1}(\cdot)+ \nu f_m(\cdot)$
		\ENDFOR
	\end{algorithmic}
\end{algorithm}

In order that we can do estimation and prediction in a computationally feasible manner on large data sets in practice, we need to apply some approximations for both the linear Gaussian process and the tree-boosted Gaussian process models. In the following, we describe these approximations. First, the integral in \eqref{marg_prob} cannot be calculated in closed form, and we approximate it using a Laplace approximation. Laplace approximations are computationally efficient, have asymptotic convergence guarantees, and are accurate for large data sets; see, e.g., \citet{kundig2024iterative}. Furthermore, to ensure that computations with Gaussian processes scale to large data sets, we use Vecchia approximations \citep{vecchia1988estimation, datta2016hierarchical, katzfuss2017general} for the latent Gaussian process $b$. In spatial statistics, Vecchia approximations have recently ``emerged as a leader among the sea of approximations" \citep{guinness2019gaussian} and are often considered as ``the most promising class of approximations" \citep{kang2023correlation, rambelli2025accuracy}. In brief, Vecchia approximations calculate an approximate sparse reverse Cholesky factor of the precision matrix of the latent Gaussian process $b$ by using an ordered conditional approximation for the density of the Gaussian process. In doing so, every row of this sparse Cholesky factor contains maximally $\tilde{m}$ non-zero entries. For prediction, a Vecchia approximation is applied to the joint distribution of a latent Gaussian process at the training and prediction points. Posterior predictive distributions for the latent Gaussian process are then obtained as conditional distributions of this approximated joint distribution. Further, predictive probabilities for the observable response variables are calculated by numerically integrating over the posterior predictive distribution of the latent Gaussian variable using, e.g., adaptive Gauss-Hermite quadrature \citep{liu1994note}. We describe the construction of Vecchia approximations for the latent Gaussian process in more detail in Appendix \ref{appendix:Vecchia}. For more information on Vecchia-Laplace approximations, including estimation and prediction, see \citet{sigrist2023latent} and \citet{kundig2024iterative}. Additionally, for increased computational efficiency, we use the iterative methods of \citet{kundig2024iterative} for estimation and prediction with Vecchia-Laplace approximations instead of Cholesky decompositions.
 
\section{Application to mortgage credit risk data}\label{section:application}
In the following, we apply our proposed models to a large U.S. mortgage credit risk data set. 

\subsection{Data and default definition}
We consider mortgage data from Freddie Mac's publicly available single-family loan-level data set\footnote{https://www.freddiemac.com/research/datasets/sf-loanlevel-dataset} from release 37. Freddie Mac provides monthly loan-level credit performance records on all mortgages that Freddie Mac purchased or guaranteed from 1999 onwards. In addition, Freddie Mac provides for every year random subsamples of 50'000 loans that originated in the corresponding year. For our analysis, we consider the union of all fully amortizing 30-year fixed-rate mortgages in all random subsets of Freddie Mac from 1999 through 2022. We model the data at a yearly frequency, and for every year, our data set contains all active mortgages that have not been terminated at the beginning of the year and that have originated before this year in one of the subsamples of Freddie Mac. A mortgage is considered as terminated when the loan defaults, reaches maturity, or when its balance is reduced to zero for example due to prepayment or sale to a third party. Furthermore, a loan is considered to be in default if it is at least 90 days delinquent. For each year and all active mortgages, we construct a default indicator that equals one if a mortgage defaults during the year and zero otherwise. We do not consider a mortgage to be active at the beginning of the year if the most recent monthly performance record is older than six months or if no performance record is available for the first three months after the loan's origination date. In addition, we restrict our sample to loans on properties located in the contiguous United States.

\begin{figure}[ht!]
\centering
   \includegraphics[width=0.9\linewidth]{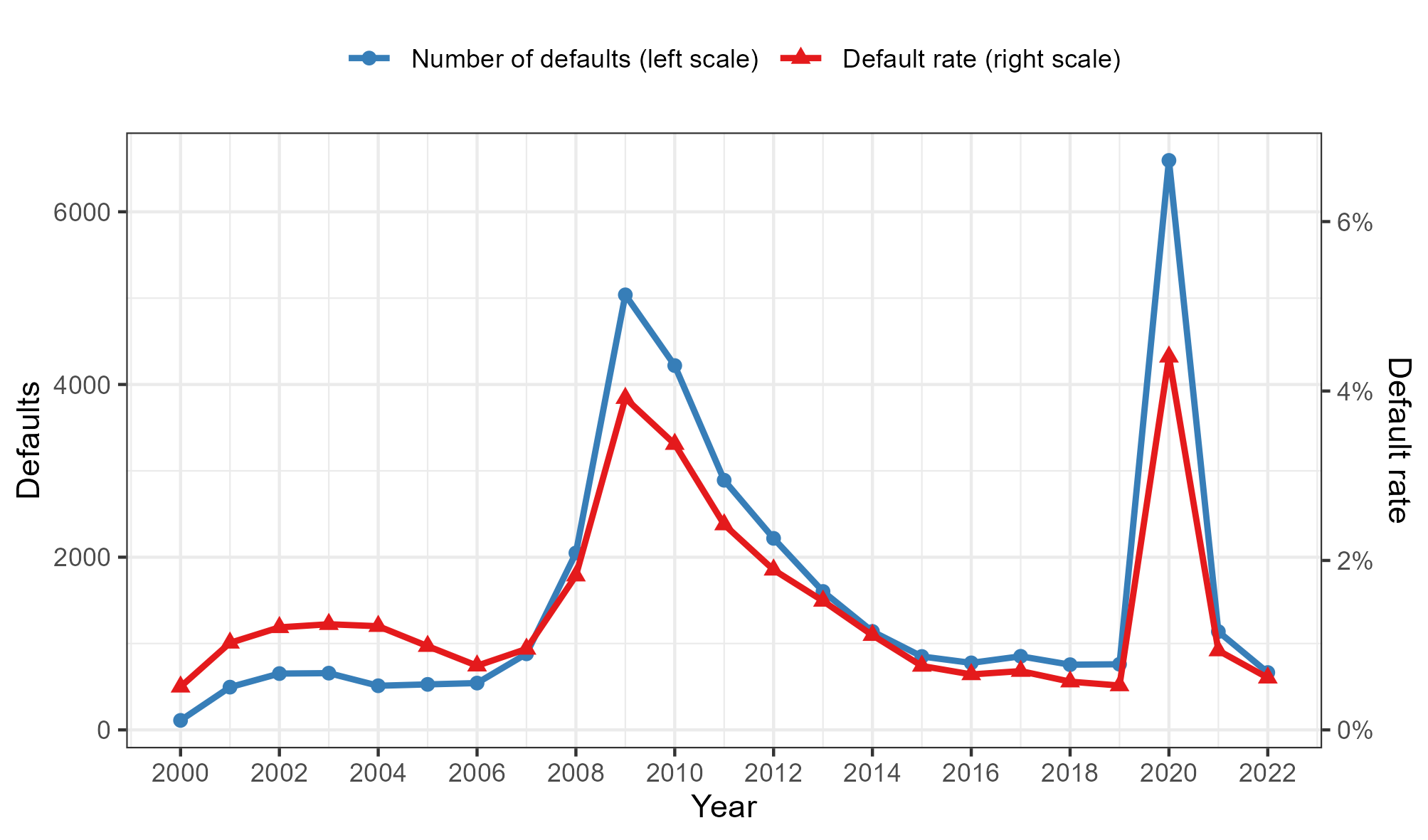}
    \caption{Number of defaults and default rate over time.}
    \label{fig:yearly_defaults}
\end{figure}

In total, our data set contains 2'256'528 loan-year observations for 538'942 different mortgages, of which 35'923 loans defaulted. Figure \ref{fig:yearly_defaults} shows the number of defaults and the default rate over time. Many defaults occur around the year 2008 during the global financial crisis, and a peculiar one-year spike occurs in the year 2020 during the COVID-19 pandemic.

\begin{figure}[ht!]
\centering
   \includegraphics[width=0.9\linewidth]{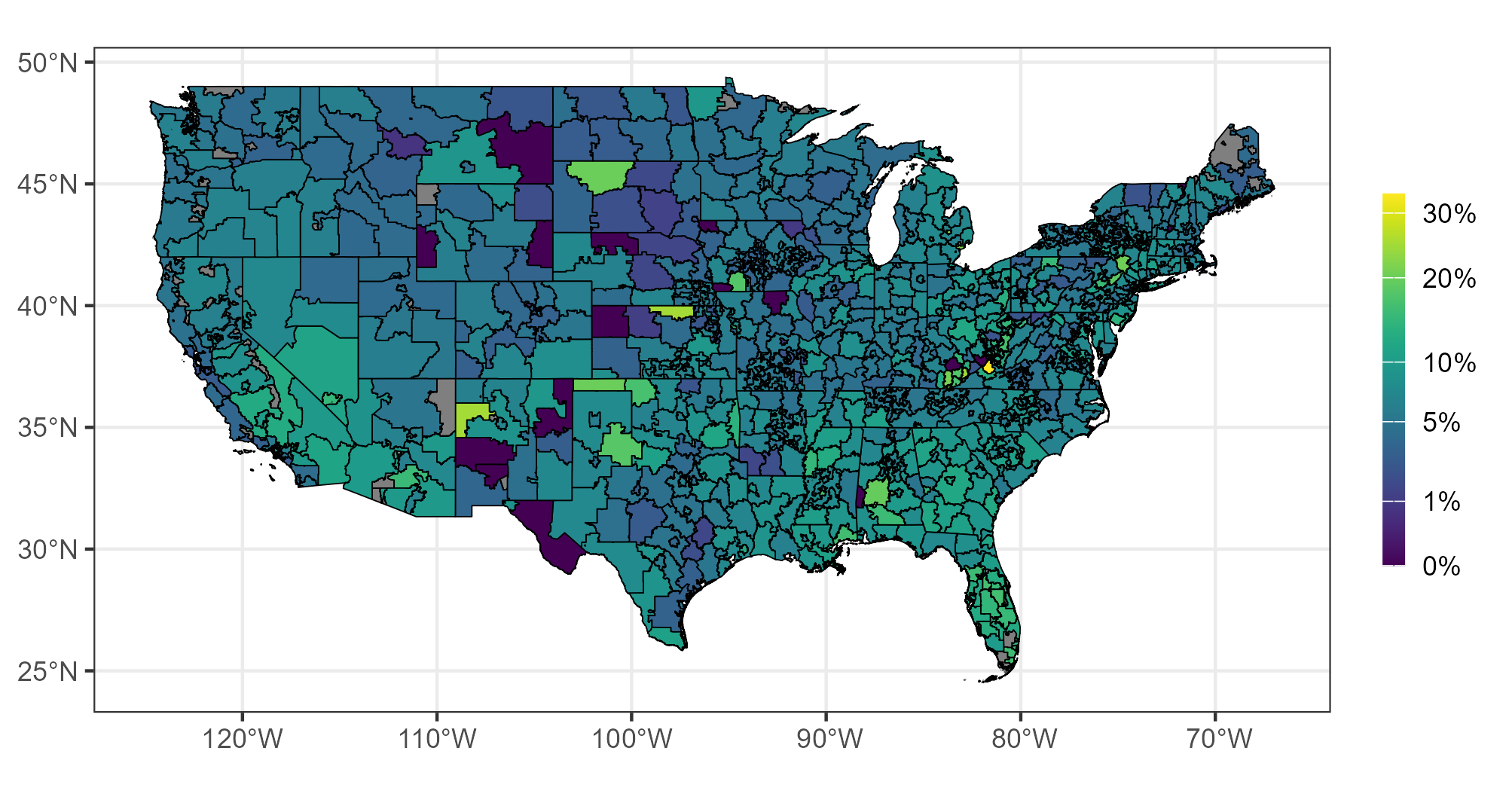}
    \caption{Spatial default rates. No data is available for the gray areas.}
    \label{fig:areal_default_rate}
\end{figure}

For the single-family loan-level data set, Freddie Mac provides the first three digits of a five-digit postal code for the property of each mortgage. For confidentiality reasons, exact locations are not available. Every three-digit postal code is associated with a specific area in the United States, and we assign the centroid coordinates of the corresponding area to every mortgage. Our data set contains mortgages from a total of 875 different areas. Figure \ref{fig:areal_default_rate} shows the aggregate default rates for the three-digit postal code areas over the years 2000 to 2022. Spatial patterns are visible such as higher default rates in the states of California, Florida, Nevada, and Arizona, and lower default rates in states such as Idaho, Wyoming, North Dakota, and Nebraska. In Figure \ref{fig:yearly_areal_default_rate} in Appendix \ref{appendix:default_rates}, we additionally report the spatial default rates separately for every year. Spatial default patterns change visibly over time.


\subsection{Predictor variables}
From the data provided by Freddie Mac, we construct several static and temporally varying predictor variables. The latter predictor variables are based on information available at the beginning of each year and include the age of the loan in months, the current loan-to-value ratio, and the interest rate spread. The current loan-to-value ratio is calculated by dividing the current unpaid principal balance by the appraised value of the property at the time of origination. To calculate the interest rate spread, we follow \cite{calabrese2024impacts} and subtract from the loan interest rates the average interest rate currently issued by Freddie Mac for 30-year fixed-rate mortgages. This average interest rate is retrieved from the Federal Reserve Bank of St. Louis\footnote{https://fred.stlouisfed.org/series/MORTGAGE30US}. Static predictor variables include the borrower's credit score at the time when the mortgage originated, additional financial indicators, and several characteristics about the property and the mortgage. These predictor variables are commonly chosen when Freddie Mac’s loan-level data set is used in the credit risk literature; see, e.g., \citet{hu2019joint} and \citet{medina2023joint}. In addition, we include a number of macroeconomic predictor variables at state and national levels, which vary over time and are primarily sourced from the U.S. Bureau of Economic Analysis\footnote{https://bea.gov}. A description of all predictor variables is provided in Table \ref{table:covariates}. Missing values for categorical predictor variables are imputed using the most frequent category, and numeric predictor variables are imputed using the mean. Summary statistics of the numeric and categorical variables are provided in Tables \ref{table:summary_numeric} and \ref{table:summary_cat}, respectively, in Appendix \ref{appendix:summary}. For all linear models, we additionally include an intercept term in the predictor variables $X_{ki}$.
\begin{table}[ht!]
    \centering
    \begin{tabular}{l p{9cm}}
        \hline
        Variable & Description\\
        \hline
        credit\_score
            & Borrower’s credit score\\ 
        longitude
            & Longitude of the centroid\\
        latitude
            & Latitude of the centroid\\
        occupancy
            & Indicates whether the property is owner-occupied (P), a second home (S), or an investment property (I).\\
        nr\_units
            & Indicates whether the property has 1, 2, 3, or 4 units.\\
        loan\_purpose
            & Indicates whether the mortgage is for cash-out refinance (C), no cash-out refinance (N), or for purchase (P).\\
        first\_time\_homebuyer
            & Indicates whether the borrower has not owned any residential property in the three years prior to the purchase of the mortgaged property.\\
        msa 
            & Indicates whether the property is reported to be located in a metropolitan statistical area or not.\\
        insurance\_percent
            & Percentage of loss coverage on the loan that an insurer is providing in the event of a loan default.\\
        orig\_dept\_to\_income
            & Original dept-to-income: borrower's monthly debt payments divided by the borrower's monthly income\\
        orig\_loan\_to\_value
            & Original combined loan-to-value: original mortgage loan plus any secondary mortgage loan divided by the mortgage appraised value\\
        orig\_unpaid\_balance
            & Original unpaid principal balance of the mortgage\\
        multiple\_borrowers
            & Indicates whether more than one borrower is obligated to repay.\\
        curr\_loan\_to\_value
            & Current loan-to-value: current unpaid principal balance divided by the mortgage appraised value\\
        ir\_spread
            & Interest rate spread\\
        n\_months
            & Age of the loan in months\\
        gdp\_growth
            & Annual gross domestic product (GDP) growth rate by state\\
        income\_per\_capita
            & Logarithmic per capita personal income by state\\
        expend\_per\_capita
            & Logarithmic per capita personal consumption expenditures by state\\
        unemployment\_rate
            & Annual unemployment rate by state\\
        hpi\_growth
            & Annual house price index growth rate by state\\
        construction\_growth
            & Annual growth rate of GDP attributable to the construction industry by state\\
        op\_surplus\_growth
            & Annual growth rate of gross operating surplus across all industries by state\\
        inflation\_rate
            & Annual U.S. inflation rate\\
        djia\_growth
            & Annual growth rate of the year-end Dow Jones Industrial Average (DJIA)\\
        \hline
    \end{tabular}
    \caption{Description of predictor variables.}
    \label{table:covariates}
\end{table}

\subsection{Models considered and implementation details}\label{model_details}
We consider the following models with increasing levels of complexity: an independent linear hazard model, a linear spatial model, a linear spatio-temporal model, a tree-boosted spatial frailty model, and a tree-boosted spatio-temporal frailty model; see Section \ref{section:models}. We have also tried using an independent linear hazard model with dummy-coded fixed effects for the three-digit postal codes. However, the estimation of this model crashed due to memory overload with the \texttt{R} package \texttt{stats} and numerical instabilities with the \texttt{R} package \texttt{biglm}. For all models that include a spatial or spatio-temporal Gaussian process, we use a Matérn covariance function as defined in Equations \eqref{matern_spatial} and \eqref{matern_spatio_temp} with smoothness parameter $\nu=1.5$ and $m=20$ nearest neighbors for estimation and prediction with Vecchia approximations. We do not estimate the smoothness parameter $\nu$ since ``binary data contain no information about [$\nu$]" \citep{de2000bayesian}. For purely spatial processes, we use the Euclidean distance to determine the nearest neighbors in Vecchia approximations and a random ordering of the spatial coordinates since this gives accurate approximations \citep{guinness2018permutation}. For the spatio-temporal models, we use a correlation-based approach to determine nearest neighbors as in \citet{kang2023correlation} and an increasing temporal order and a random spatial order for coordinates with the same time. Following \citet{kang2023correlation}, the nearest neighbors for the spatio-temporal models are redetermined in every iteration that is a power of two whenever an optimization algorithm is used for determining the parameters $\theta$. For prediction with Vecchia approximations, the observed points appear first in the ordering, and the Gaussian process at prediction points is only conditioned on the training data in the Vecchia approximation. For finding optima for the parameters $\theta$, we use the limited-memory BFGS algorithm. Estimation and prediction with the linear spatial, linear spatio-temporal, tree-boosted spatial, and tree-boosted spatio-temporal frailty models is done using the \texttt{GPBoost}\footnote{https://github.com/fabsig/GPBoost} library version 1.5.8 \citep{sigrist2021gpboost}. The code for reproducing the mortgage credit risk application and generating the data set is publicly available at \url{https://github.com/pkuendig/SpaceTimeML}.

\subsection{Sample split for model evaluation and choosing tuning parameters}
All the following results are based on temporal out-of-sample predictions. Specifically, we conduct one-year-ahead default predictions for each year starting from 2008 through 2022 using an expanding window training data approach. In other words, we first consider all loan-year observations up to and including the year 2007 as training data and make predictions for all loan-year observations of 2008. We then continue by expanding the training window by one year and using the subsequent year as test data. Under this train-test split, the test data sets contain both new loans, which originated in the most recent year before the test data time period and are therefore not included in the training data, and older but still active loans, for which the training data contains historical observations.

For choosing tuning parameters, we analogously split every training window data set into an inner training data set and a validation data set. The validation data contains all loan-year observations of the most recent year in the training data, and the inner training data consists of all samples excluding this most recent year. Tuning parameters are chosen by estimating models on the inner training data and selecting the combinations of tuning parameters that maximize the area under the receiver operating characteristic curve (AUC) on the validation data. The candidate tuning parameters for the tree-boosted spatial and spatio-temporal frailty models are shown in Table \ref{table:candidatesLaGaBoost} in Appendix \ref{tune_pars}.

\subsection{Prediction of individual default probabilities}
We first apply the above-described models to predict one-year-ahead default probabilities of individual mortgage loans for every year from 2008 through 2022. The predictive default probabilities are evaluated using the following prediction accuracy measures: the AUC, the H-measure, the average log-loss, the Brier score, and the expected calibration error (ECE). The AUC measures predictive discrimination ability and can be interpreted as the probability that the predictive probability for a randomly drawn default event is higher than the predictive probability for a randomly drawn loan-year observation without a default. The ECE assesses calibration. A predictive probability $p$ is calibrated if the corresponding event (=default) occurs in $100 \times p$ percent. We determine the boundaries of the bins for the ECE by using $20$ equally-spaced quantiles of the empirical distribution of all predictive probabilities of all models and years pooled together. The H-measure, the log-loss, and the Brier score measure overall predictive accuracy. See, e.g., \citet{dimitriadis2023evaluating} for more information on these prediction accuracy measures.

Table \ref{table:prob_results} reports the AUC, H-measure, mean log-loss, Brier score, and ECE for every model, averaged over the 15 years for which we perform one-year-ahead default predictions. We find that the predictive default probabilities of the tree-boosted spatio-temporal frailty model are the most accurate in terms of AUC and H-measure, while predictive default probabilities of the purely spatial LaGaBoost model are the most accurate in terms of mean log-loss, Brier score, and ECE. Adding a spatial frailty variable to an independent linear hazard model improves the prediction accuracy. Compared to a purely spatial linear model, the linear spatio-temporal model has a higher prediction accuracy in terms of AUC and H-measure, but a lower accuracy in terms of mean log-loss, Brier score, and ECE. Overall, predictive default probabilities obtained using the tree-boosted spatial model are consistently more accurate than those obtained using the linear spatial model. The same holds true when comparing the tree-boosted spatio-temporal model and the linear spatio-temporal model. This suggests the presence of non-linearities and/or interactions among the predictor variables, a conclusion we corroborate through a more detailed analysis of the estimated models in Section \ref{interpret}.

\begin{table}[ht]
    \centering
    \begin{tabular}{llllll}
        \hline
            & AUC & H-measure & Mean log-loss & Brier score & ECE \\ 
        \hline
            Linear independent & 0.7676 & 0.2024 & $7.926 \times 10^{-2}$ & $1.663 \times 10^{-2}$ & $1.128 \times 10^{-2}$ \\ 
            Linear spatial & 0.7702 & 0.2050 & $7.924 \times 10^{-2}$ & $1.661 \times 10^{-2}$ & $1.112 \times 10^{-2}$ \\ 
            Linear spatio-temporal & 0.7711 & 0.2081 & $8.124 \times 10^{-2}$ & $1.697 \times 10^{-2}$ & $1.342 \times 10^{-2}$ \\ 
            LaGaBoost spatial & 0.7716 & 0.2065 & $\mathbf{7.816 \times 10^{-2}}$ & $\mathbf{1.633 \times 10^{-2}}$ & $\mathbf{8.993 \times 10^{-3}}$ \\ 
            LaGaBoost spatio-temporal & \textbf{0.7772} & \textbf{0.2149} & $7.944 \times 10^{-2}$ & $1.658 \times 10^{-2}$ & $1.066 \times 10^{-2}$ \\ 
        \hline
    \end{tabular}
    \caption{AUC, H-measure, average log-loss, Brier score, and ECE averaged over the 15 years for which one-year-ahead default predictions are calculated.}
    \label{table:prob_results}
\end{table}

\begin{figure}[ht!]
\centering
    \includegraphics[width=0.9\linewidth]{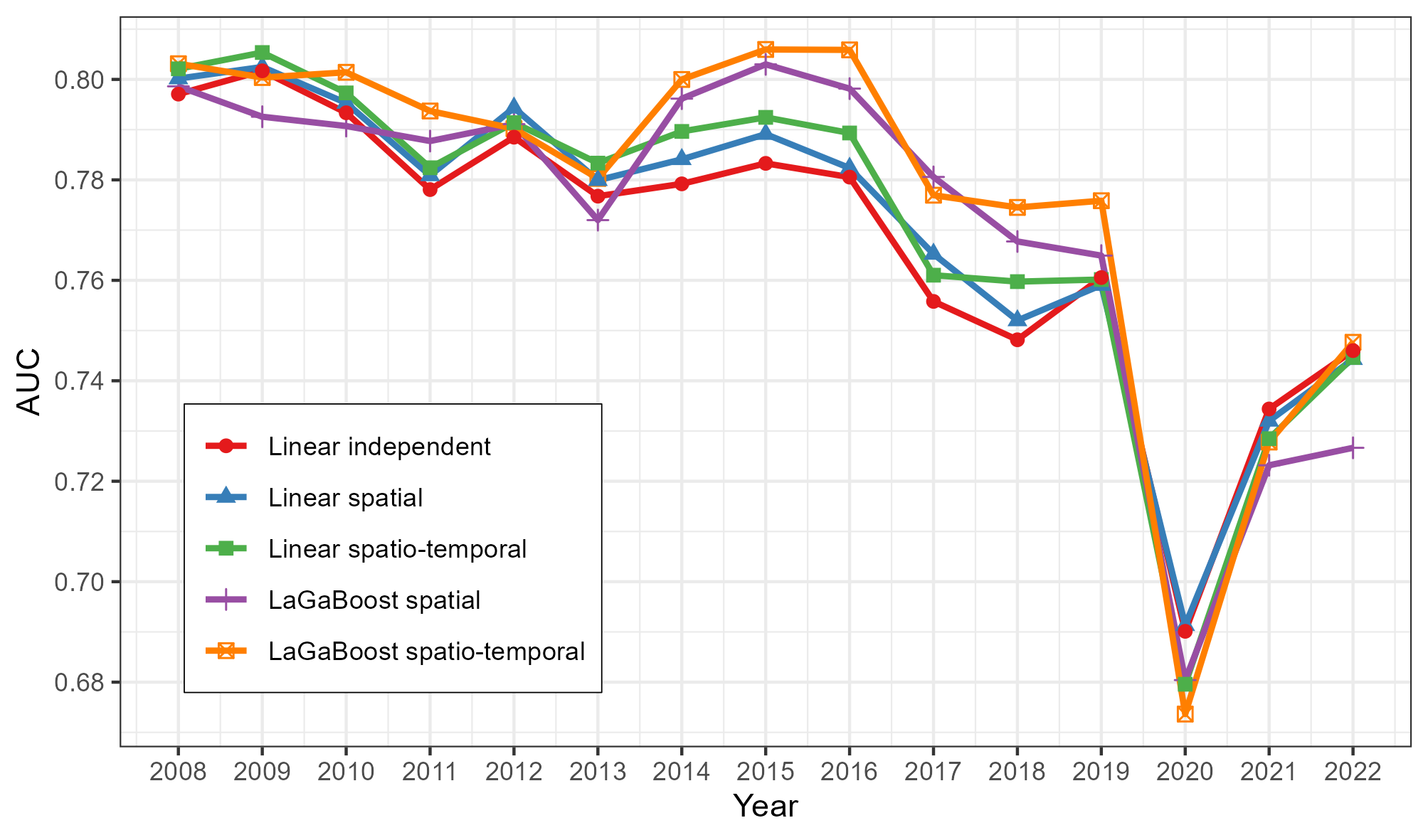}
    \caption{Temporal out-of-sample test area under the receiver operating characteristic curve (AUC) (higher = better).}
    \label{fig:auc}
\end{figure}

Figure \ref{fig:auc} additionally reports the AUC of every model over time. We observe that the AUCs are relatively high for all models and for all years before the COVID-19 pandemic. In particular, the AUC remains at high levels during the global financial crisis around the year 2008. However, in the year 2020 during the COVID-19 pandemic, the AUC drops to considerably lower levels for all models and increases again in the subsequent years. Similar patterns over time are observed for the H-measure and the ECE; see Figures \ref{fig:H_measure} and \ref{fig:ECE} in Appendix \ref{add_res_ind}. We interpret this sudden decrease in prediction accuracy of all models in 2020 in the sense that an artificial, external shock in the form of lock-downs led to very different default mechanisms compared to normal times and also compared to the global financial crisis around the year 2008. 

It is important to note that although the observed differences in AUC between the models are relatively small, even small improvements in prediction accuracy yield large economic gains, as the value of the mortgage portfolio of all active loans is in the billions of US dollars. Furthermore, we report in Table \ref{table:DeLong} in Appendix \ref{add_res_ind} p-values of pairwise DeLong tests \citep{delong1988comparing} to assess whether the differences in AUCs of the pooled predictive default probabilities over all test years are significant across the different models. We find that all models have significantly different AUCs at the 5\% level, except when comparing the linear independent and linear spatial models.

In Table \ref{table:runtimes} in Appendix \ref{appendix:runtimes}, we also report the average wall clock times for training the different models on data up to the year 2014. The spatio-temporal models require more time compared to the spatial models due to the higher dimensionality of the space-time Gaussian processes. Furthermore, the tree-boosted frailty models are slower compared to their corresponding linear counterparts. In detail, the estimation times for the spatial and spatio-temporal tree-boosting models are approximately one and 30 minutes, respectively. Calculations were performed on a laptop with an Intel i7-12800H processor and 32 GB of random-access memory.

\subsection{Prediction of loan portfolio loss distributions}
In the following, we apply the different models for predicting one-year-ahead loss distributions of annual mortgage portfolios containing all active loans from the beginning of every year. If a loan defaults, this results in a loss corresponding to the loan's unpaid principal balance at the time when predictions are made. Predictive loss distributions of portfolios are approximated by simulating 100'000 times sums of Bernoulli variables with according predictive default probabilities. For the models with a latent Gaussian process, we use a two-step simulation approach as follows. In every simulation run, first, a sample from the posterior predictive distribution of the latent Gaussian process is drawn which is then added to the predictions of the fixed effects function to obtain predictive probabilities which, in a second step, are used to simulate default indicator variables.

The prediction of one-year-ahead loss distributions is performed for every model and year from 2008 to 2022. Quantifying the accuracy of predictive loan portfolio loss distributions is inherently difficult due to the relatively small number of temporally independent observations. Nonetheless, we evaluate the accuracy of the entire predictive loss distributions using the continuous ranked probability score (CRPS) \citep{gneiting2007strictly}, of upper 99\% predictive quantiles using the 99\% quantile loss \citep{koenker1999goodness}, and of the mean using the root-mean-square error (RMSE). The CRPS is a proper scoring function that generalizes the absolute error to probabilistic predictions and is defined as $\text{CRPS}(F,\text{L}) =\int_{-\infty}^{\infty} (F(y) - \mathds{1}_{\{y \geq \text{L}\}})^2 \,dy$, where $F$ is a cumulative predictive loan portfolio loss distribution function and $L$ is the realized portfolio loss. Furthermore, upper tails of loss distributions are of particular interest for risk management purposes. We thus consider predictive 99\% quantiles and evaluate them using the corresponding quantile loss. The latter is given by $S(q_{\alpha};\text{L}) = (\text{L}-q_{\alpha})(\alpha - \mathds{1}_{\{\text{L} \leq q_{\alpha}\}})$, where $q_{\alpha}$ denotes the predictive $\alpha$ quantile. This asymmetric quantile loss function is a proper scoring rule \citep{gneiting2007strictly}, and it penalizes observations L which are higher than the predicted quantile $q_{\alpha}$ more heavily. The RMSE is calculated by comparing means of predictive loss distributions with realized portfolio losses.


Table \ref{table:portfolio_results} reports the CRPS, the 99\% quantile loss, and the RMSE for the 15 years used as test data.  Overall, the tree-boosted spatial and spatio-temporal frailty models are the most accurate models. In more detail, we observe that the tree-boosted spatio-temporal model has the lowest quantile loss. In other words, this model yields the most accurate predictive upper quantiles of portfolio loss distributions. Furthermore, the tree-boosted spatial model has the lowest CRPS and RMSE, i.e., this model is the most accurate for the center of portfolio loss distributions. In detail, the difference in RMSE between the tree-boosted spatial model and the linear spatial model are approximately US \$35 million. In addition, the predictive portfolio loss distributions of the linear spatial model are more accurate in terms of the CRPS and the RMSE compared to the independent linear hazard model.
\begin{table}[ht!]
    \centering
    \begin{tabular}{llll}
      \hline
        & CRPS & 99\% quantile loss & RMSE \\ 
      \hline
        Linear independent & $2.699 \times 10^{8}$ & $1.158 \times 10^{8}$ & $4.269 \times 10^{8}$ \\ 
        Linear spatial & $2.627 \times 10^{8}$ & $1.210 \times 10^{8}$ & $4.178 \times 10^{8}$ \\ 
        Linear spatio-temporal & $3.128 \times 10^{8}$ & $1.188 \times 10^{8}$ & $5.615 \times 10^{8}$ \\ 
        LaGaBoost spatial & $\mathbf{2.034 \times 10^{8}}$ & $1.072 \times 10^{8}$ & $\mathbf{3.823 \times 10^{8}}$ \\ 
        LaGaBoost spatio-temporal & $2.598 \times 10^{8}$ & $\mathbf{1.021 \times 10^{8}}$ & $4.960 \times 10^{8}$ \\ 
       \hline
    \end{tabular}
    \caption{Accuracy of one-year-ahead predictive loan portfolio loss distributions.} 
    \label{table:portfolio_results}
\end{table}

\begin{figure}[ht!]
\centering
   \includegraphics[width=0.9\linewidth]{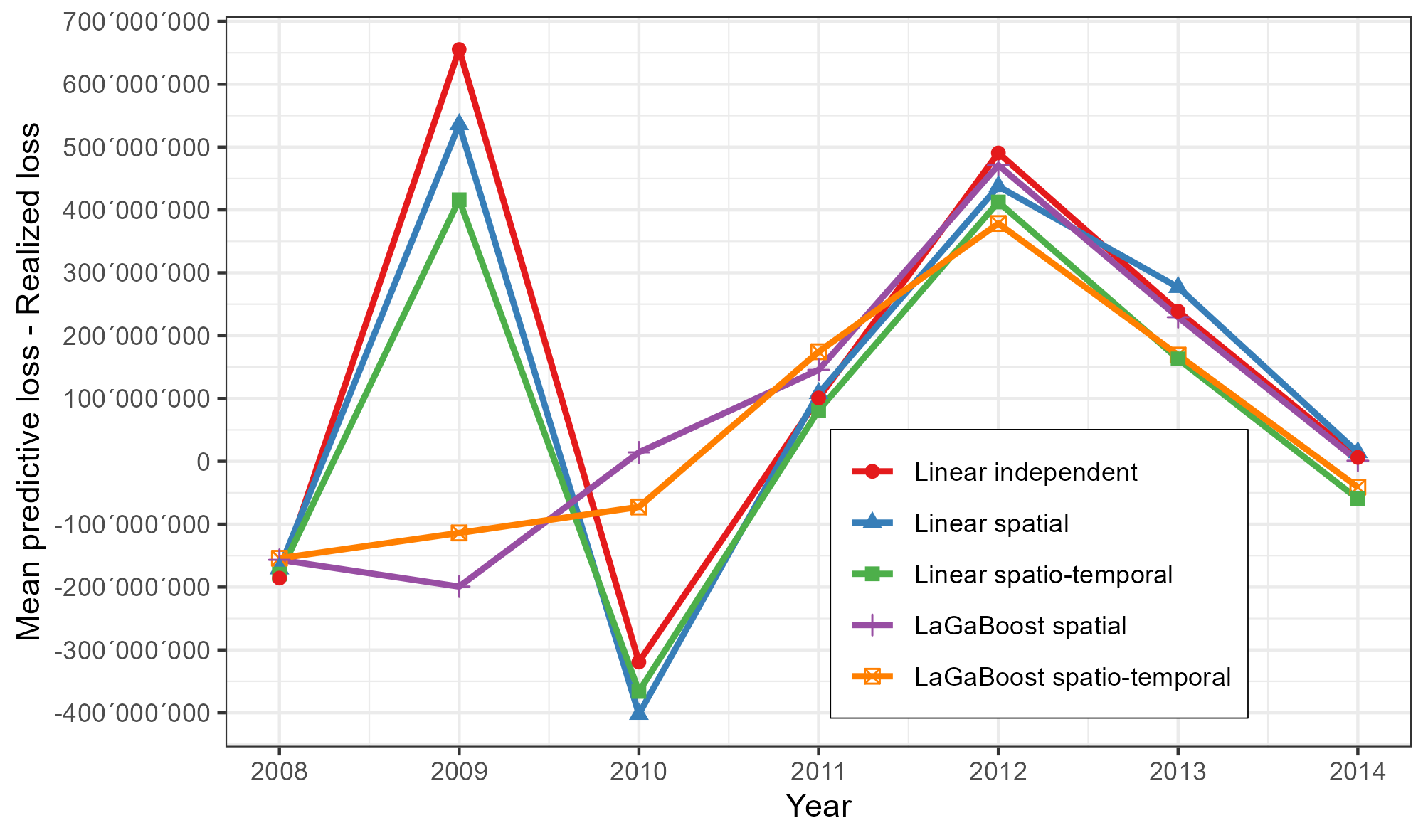}
    \caption{Differences between means of the predictive loss distributions and realized portfolio losses.}
    \label{fig:diff_loss}
\end{figure}

Next, Figure \ref{fig:diff_loss} shows the differences between the means of the predictive portfolio loss distributions and the realized portfolio losses over the years 2008 to 2014. We see that the tree-boosted spatial and spatio-temporal frailty models clearly predict the most realistic mean portfolio losses during the global financial crisis around the year 2008. The linear models clearly overestimate the portfolio loss in the second year of the global financial crisis, 2009, a trend that is reversed by underestimating it in 2010. In Figure \ref{fig:diff_loss_all} in Appendix \ref{appendix:loss_distribution}, we also show the differences between the predictive mean portfolio losses and the realized portfolio losses for all prediction years from 2008 to 2022.

\begin{figure}[ht!]
\centering
   \includegraphics[width=0.9\linewidth]{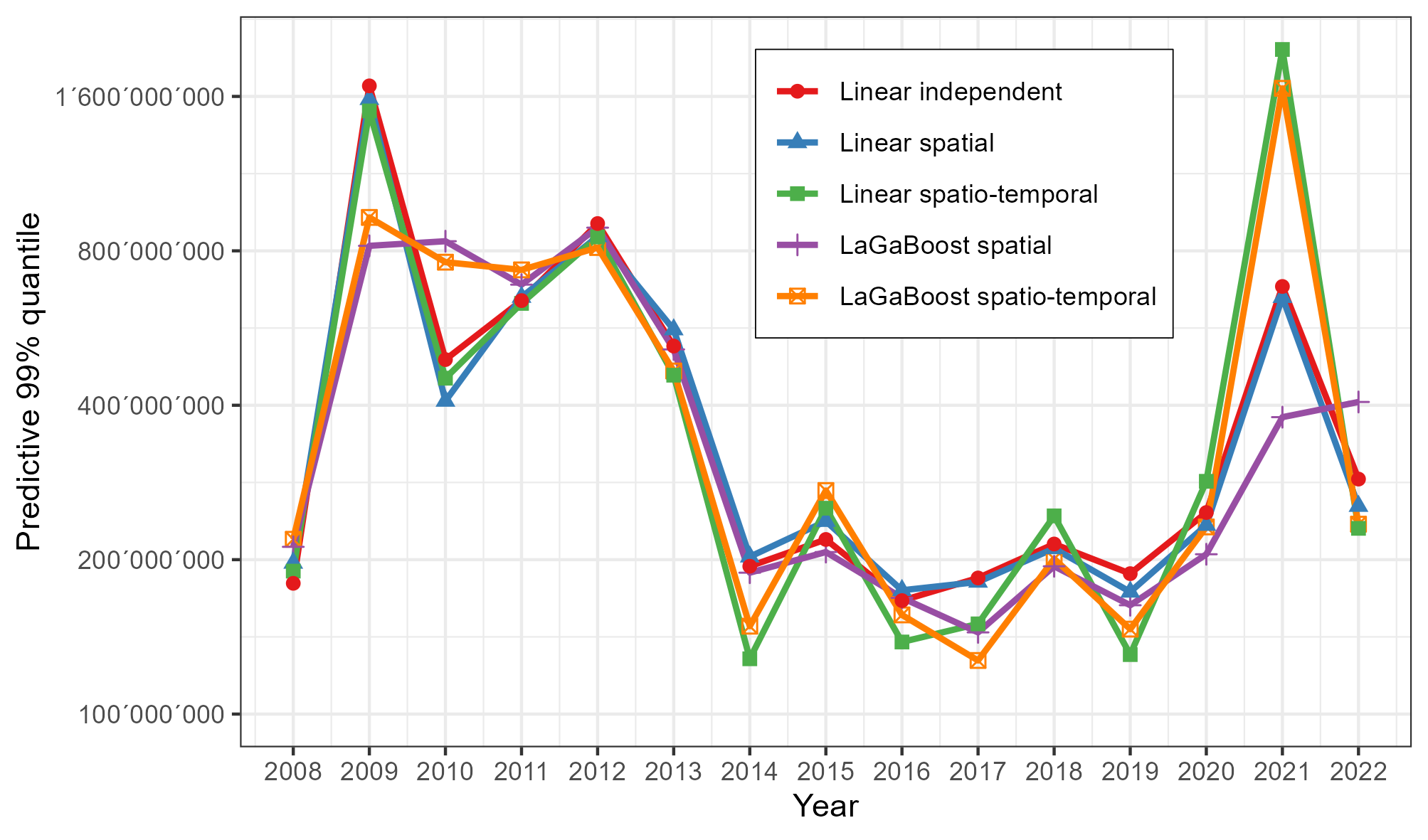}
    \caption{Predictive $99\%$ quantiles of one-year-ahead loan portfolio loss distributions.}
    \label{fig:q99}
\end{figure}

Figure \ref{fig:q99} additionally shows the time series of predictive $99\%$ quantiles of one-year-ahead portfolio loss distributions. We again find that the tree-boosted frailty models predict the most consistent upper tail loan portfolio losses at the beginning and end of the global financial crisis. Similar to the mean portfolio loss, linear models initially overestimate, then underestimate, and finally overestimate upper tail losses during the years of the financial crisis. The linear independent model without a frailty Gaussian process generates the smallest predictive $99\%$ quantiles before the global financial crisis in 2008. The sudden one-year default spike in 2020 during the COVID-19 pandemic causes all models to predict very high upper tail loan portfolio losses for the subsequent year 2021. However, the overestimation of the upper tail portfolio loss in Figure \ref{fig:q99} as well as the mean portfolio loss in Figure \ref{fig:diff_loss_all} in Appendix \ref{appendix:loss_distribution} in 2021 is particularly pronounced for the linear spatio-temporal model and the tree-boosted spatio-temporal model. Purely spatial frailty models, on the other hand, yield lower predictive quantiles for the year 2021 since their frailty components are constant over time.

\subsection{Model interpretation}\label{interpret}
\begin{figure}[ht!]
\centering
   \includegraphics[width=\linewidth]{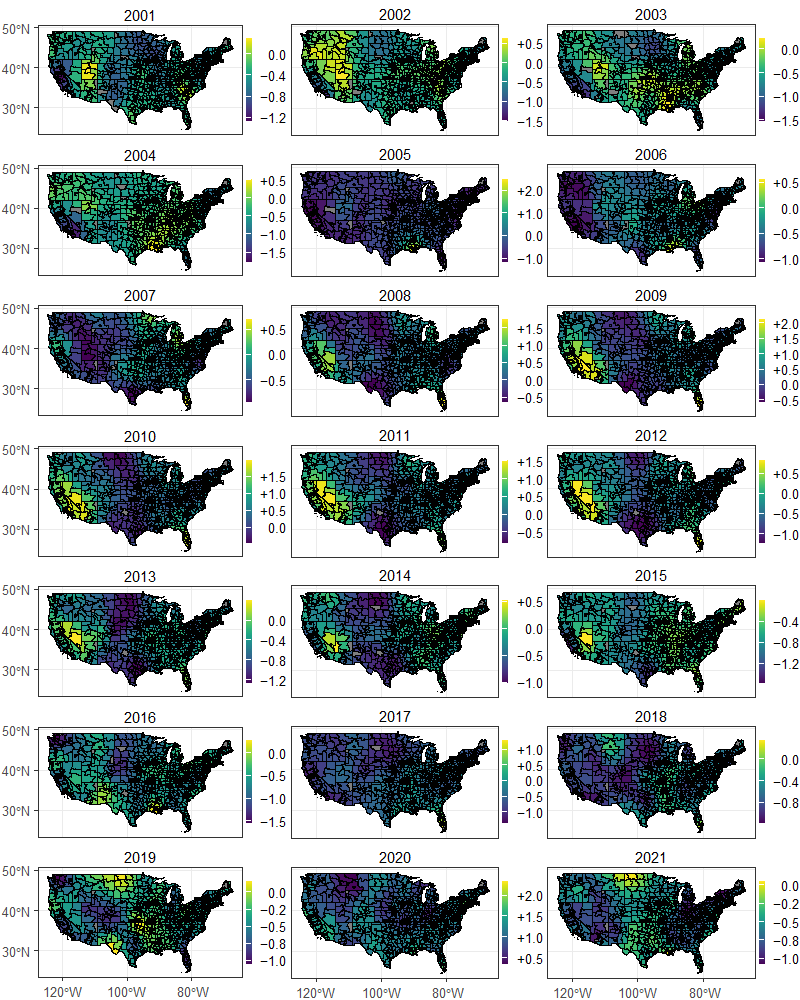}
    \caption{
    Posterior mean for the latent Gaussian process in the tree-boosted spatio-temporal frailty model when training on data up to the year 2021.}
    \label{fig:latent_b_gpboost}
\end{figure}

In the following, we aim to better understand the functioning of the tree-boosted and linear spatio-temporal frailty models. In Figure \ref{fig:latent_b_gpboost}, we show the mean of the posterior distribution of the latent spatio-temporal frailty Gaussian process for the tree-boosted spatio-temporal model when training on data up to and including the year 2021, which corresponds to the model for predicting defaults in the most recent year 2022. Due to space constraints, posterior means are not shown for the year 2000. We additionally show these posterior mean maps for the linear spatio-temporal model in Figure \ref{fig:latent_b_linear} in Appendix \ref{post_mean_lin}. We observe that there is considerable variation in the latent frailty variable over space and this variation changes over time. In particular, for the years 2005 and 2006, posterior means for the latent Gaussian variable are high in the New Orleans area, which can be explained by high default rates following Hurricane Katrina in 2005. In 2008 and subsequent years, posterior means are high in the so-called ``bubble" states California, Florida, Nevada, and Arizona, where house prices rose particularly rapidly in the run-up to the subprime mortgage crisis \citep{haughwout2011real}. For the year 2020, posterior means are high due to the generally large number of defaults, with the highest values found in the New York City area (not visible at this resolution). This is not surprising given that New York City was one of the first and most affected areas in the U.S. by the COVID-19 pandemic. 
Comparing posterior means of the tree-boosted spatio-temporal frailty model with those of the linear spatio-temporal model, we find similar patterns over space and time.

\begin{figure}[ht!]
    \centering
   \includegraphics[width=0.9\linewidth]{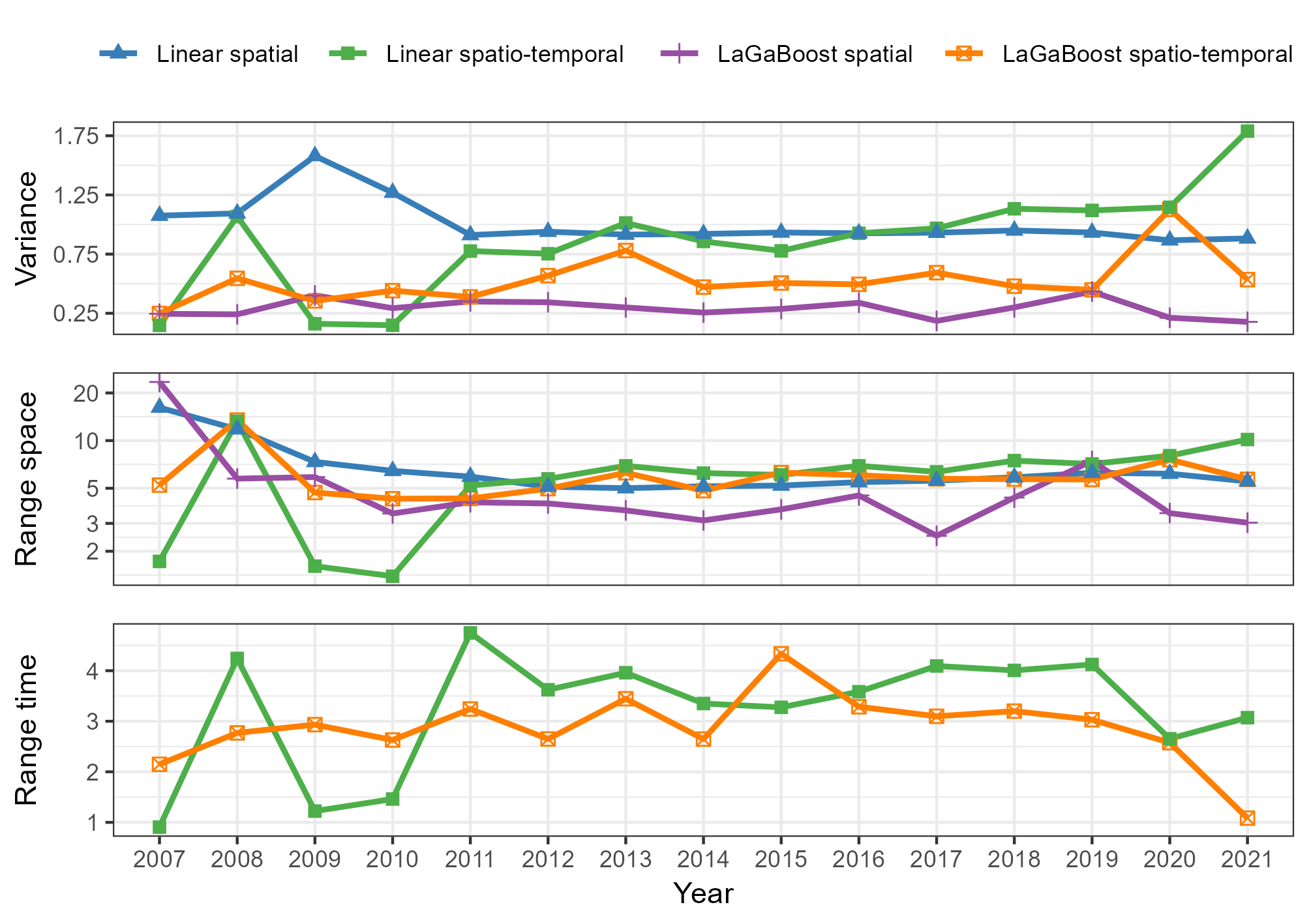}
    \caption{Estimated covariance parameters for different expanding window training data sets.}
    \label{fig:cov_pars}
\end{figure}

In Figure \ref{fig:cov_pars}, we report the estimated covariance parameters for the different expanding window training data sets for the linear spatial, the linear spatio-temporal, the tree-boosted spatial, and the tree-boosted spatio-temporal frailty models. For the latter, estimated variance and range parameters are relatively constant until the year 2020, when the variance parameter increases from 0.45 to 1.13, and in the following year 2021, the range parameter for time decreases from 2.58 to 1.08. These estimates are likely a consequence of the sudden atypical one-year default spike in 2020 during the COVID-19 pandemic. Since default mechanisms were likely very different in this year compared to the past due to an artificial external shock, less variation in the response variable is explained by the fixed effects and the latent Gaussian variable becomes more important as can be inferred from the higher marginal variance. Subsequently, defaults return to normal levels in 2021, and the estimated correlation with previous years is lower when including 2020 in the training data as can be seen from the lower estimated temporal range parameter. In both the linear spatial model and the tree-boosted spatial model, the estimated range parameters decrease during the initial training years and then remain relatively constant for the remaining years. Furthermore, the estimated variance parameter of the linear spatial model is consistently higher than the one of the tree-boosted spatial model throughout all training years. This indicates that the latent frailty variable plays a more important role in the linear spatial model, since its linear deterministic function can explain less variation in the data compared to the tree-ensemble of the LaGaBoost spatial model.
Similarly, for training years 2011 to 2019, the estimated variance parameter of the linear spatio-temporal model is also consistently higher than that of the LaGaBoost spatio-temporal model. This difference can also be attributed to the different explanatory power of their respective predictor functions.

\begin{figure}[ht!]
    \centering
   \includegraphics[width=0.9\linewidth]{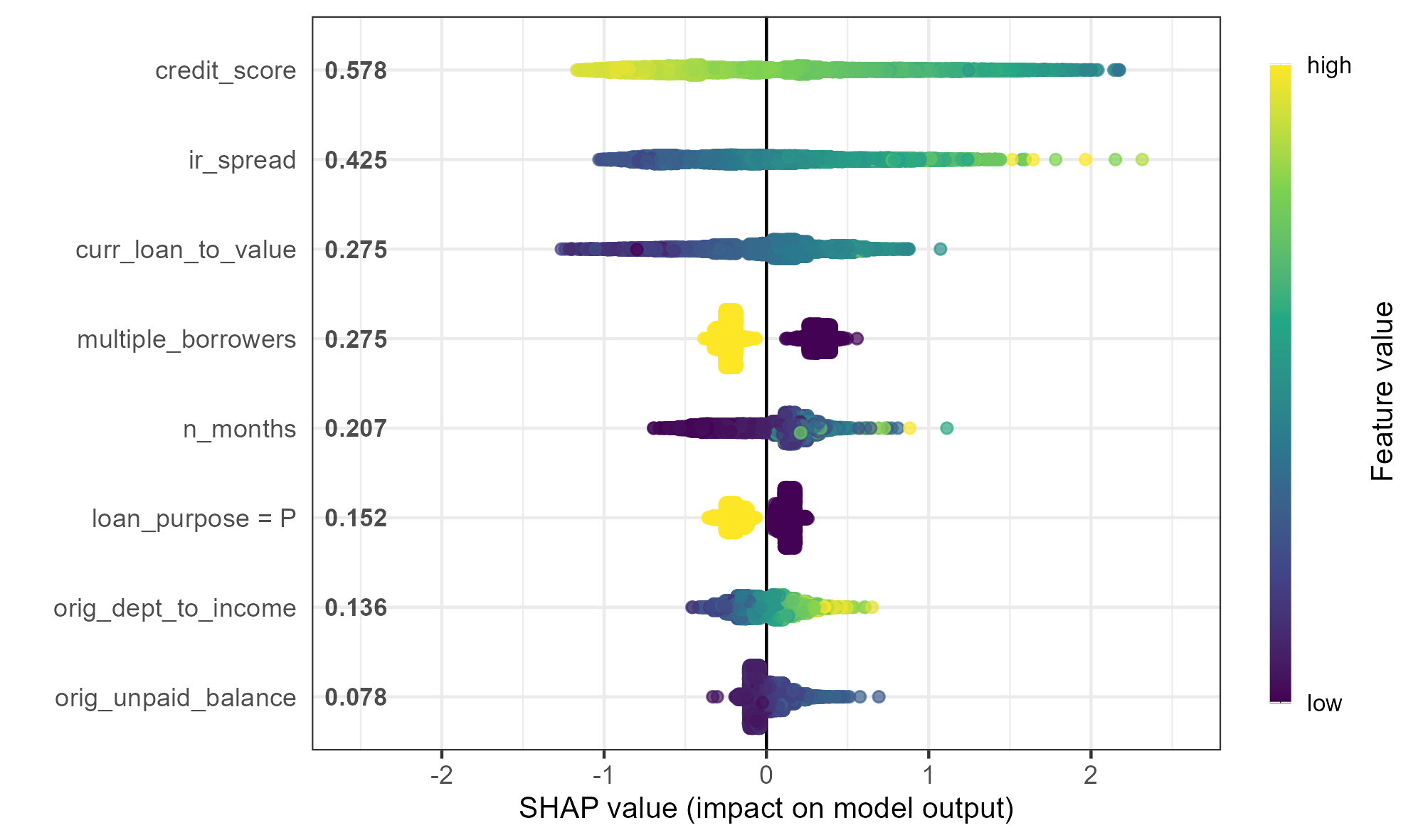}
    \caption{SHAP values for the tree-boosted spatio-temporal frailty model when training on data up to the year 2014.}
    \label{fig:shap_values}
\end{figure}

\begin{figure}[ht!]
    \centering
   \includegraphics[width=0.9\linewidth]{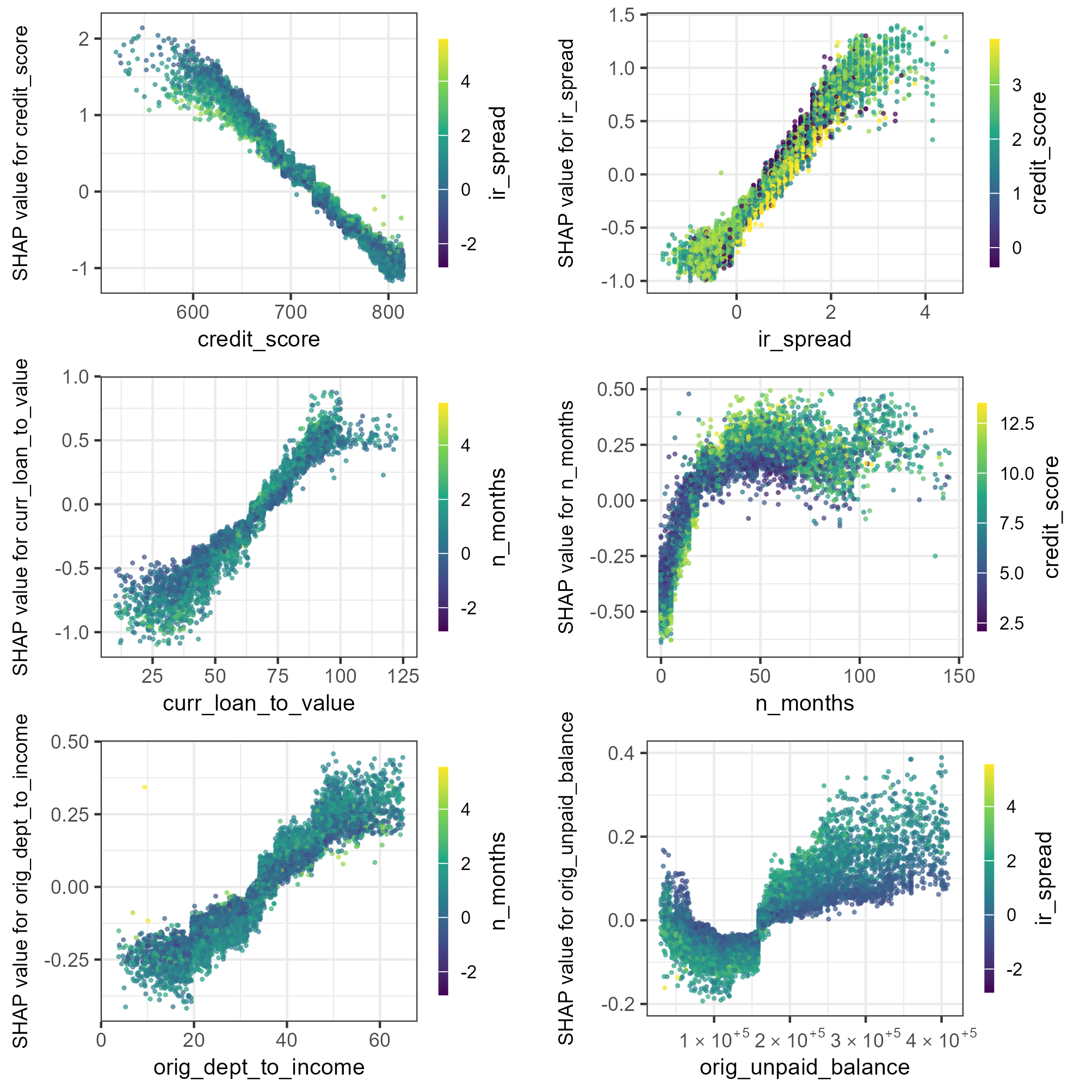}
    \caption{SHAP dependence plots for the tree-boosted spatio-temporal frailty model when training on data up to the year 2014.}
    \label{fig:shap_dependence}
\end{figure}

For understanding the function $F(\cdot)$ of the tree-boosted frailty models, we use SHAP values and SHAP dependence plots \citep{lundberg2017unified}. Specifically, we consider SHAP values for the tree-boosted spatio-temporal model trained on data up to the year 2014. We chose this training window because, for the corresponding prediction year 2015, there are pronounced differences in the accuracy of predictive default probabilities between the tree-boosted and linear frailty models. SHAP values are calculated using 10'000 randomly selected instances of the training data. This subsampling is done to reduce the computational complexity and to avoid that SHAP dependence plots are overcrowded. Using a larger random subsample or a different subsample gives almost identical results (results not shown). Figure \ref{fig:shap_values} shows the SHAP values. The predictor variables are ordered according to the average of the absolute values of the SHAP values. According to these results, the eight most important predictor variables in descending order are the credit score (credit\_score), the interest rate spread (ir\_spread), the current loan-to-value (curr\_loan\_to\_value), the indicator whether multiple borrowers are obligated to repay (multiple\_borrowers), the age of the loan (n\_months), the indicator whether the mortgage is used for purchase (loan\_purpose = P), the original debt-to-income (orig\_dept\_to\_income), and the original unpaid principal balance (orig\_unpaid\_balance). For the six numeric variables thereof, we report the corresponding SHAP dependence plots in Figure \ref{fig:shap_dependence}. In all dependence plots, we observe strong interaction effects. This can be inferred from the large vertical scatter of the SHAP values and the systematic colored relationship with other predictor variables. For example, the slope of the SHAP values for the credit score variable (credit\_score) is less negative if the interest rate spread variable (ir\_spread) is positive. I.e., when interest rate spreads are high, the individual credit score matters less compared to when interest rate spreads are low. Furthermore, we find strong non-linear effects for the age of the loan (n\_months) and the original unpaid principal balance (orig\_unpaid\_balance). For instance, the age of a loan is positively related to the default probability up to a certain age of approximately three years, and then the effect flattens out. In addition, the variables interest rate spread (ir\_spread), current loan-to-value (curr\_loan\_to\_value), and original debt-to-income (orig\_dept\_to\_income) also have clearly non-linear relationships with slopes that change markedly for large and small values of these variables. For instance, increasing interest rate spreads are related to higher default probabilities, but this effect only holds up to a certain level of approximately two and a half percent, after which the relationship levels off, and additional increases in interest rate spreads are not associated with higher default probabilities. Furthermore, the debt-to-income (orig\_dept\_to\_income) appears to have an approximately logistic-shaped effect on the default probability with the relationship being almost flat for both small debt-to-income values below 20 percent and large values above 50 percent, and in between, the effect of debt-to-income is approximately linear with a slope that interacts, among other things, with the loan age. We conclude that there are strong interaction and non-linear effects. This is likely also the reason why the tree-boosted spatial and spatio-temporal frailty models achieve higher prediction accuracy than their respective linear counterparts.

\section{Conclusion} \label{section:conclusion}
We introduce a novel machine learning model combining tree-boosting with a latent spatio-temporal Gaussian process that accounts for frailty correlation. We compare our proposed model to an independent linear hazard model, a linear model with spatial frailty effects, and a linear model with spatio-temporal frailty effects and find that predictive default probabilities and predictive loan portfolio loss distributions obtained with our novel model are more accurate. Although the differences in prediction accuracy for individual default probabilities are relatively small, they are nonetheless important, as even small differences are relevant for large portfolios of mortgages. For instance, during the global financial crisis around the year 2008, we observe large differences in predictive loan portfolio loss distributions with mean losses of frailty machine learning models being almost US \$1 billion more accurate compared to linear models. Using interpretability tools for machine learning models, we find that there are strong interactions and non-linear effects in the predictor variables. This is further evidence that linear models likely assume a misspecified functional form. In addition, we find that there are sizable spatio-temporal frailty effects.

A potential limitation of the spatio-temporal models we considered in this article is that they overestimate the portfolio loss in the year 2021 after the COVID-19 pandemic which lead to a rather peculiar one-year peak in default probabilities during the year 2020. On the other hand, linear and machine learning models that only account for spatial frailty do not have this drawback. Future research can investigate whether and how such an overestimation can be mitigated, for instance, using other forms of spatio-temporal models. In general, future research can systematically analyze and compare different space-time covariance functions for spatio-temporal frailty modeling.

For confidentiality reasons, the exact locations of the mortgaged properties are not included in the single-family loan-level data set from Freddie Mac we used in this article. Consequently, spatial and spatio-temporal frailty correlation could only be modeled at a relatively coarse resolution. Having exact coordinates could potentially further improve the accuracy of spatial and spatio-temporal frailty models. Future research is required to investigate whether modeling of spatial and spatio-temporal frailty correlations using more granular credit risk data sets leads to more accurate predictions. We expect potential improvements not just for mortgages but in general when modeling loans associated with a specific spatial location such as SME, credit card, student, and peer-to-peer loans. 

\section*{Acknowledgments}
We would like to thank Raffaella Calabrese for her helpful advice in gathering the mortgage data, Pascal Gantenbein and Christian Kleiber for their useful feedback, and three anonymous reviewers for their constructive comments, which helped to improve this article. This research was partially supported by the Swiss Innovation Agency - Innosuisse (grant number `55463.1 IP-ICT'). 

\bibliographystyle{abbrvnat}
\bibliography{bib_default_space_time.bib}
\clearpage
\appendix

\section{Appendix}

\subsection{Vecchia approximations for the latent Gaussian process} \label{appendix:Vecchia}
We apply a Vecchia approximation to the distribution of the latent Gaussian variable $b \in \mathbb{R}^R$:
\begin{equation*}
\begin{split}
    p(b|\theta) &= \prod_{r=1}^R p(b_r|(b_1,\dots,b_{r-1}),\theta)\\
    & \approx \prod_{r=1}^R p(b_r|b_{N(r)},\theta), 
\end{split}
\end{equation*}
where $b_{N(r)}$ are subsets of the conditioning sets $(b_1,\dots,b_{r-1})$, and $N(r)$ denotes the corresponding subsets of indices. If $r > \tilde{m}+1$, $N(r)$ is often chosen such that it contains the indices of the $\tilde{m}$ nearest neighbors of $b_r$ in the conditioning set. Otherwise, $N(r)$ equals $(1,\dots,r-1)$. Using the properties for conditional multivariate normal distributions, we get:
\begin{equation*}
    p(b_r|b_{N(r)},\theta) = \mathcal{N}(A_r b_{N(r)},D_r),
\end{equation*}
where
\begin{equation*}
\begin{split}
    A_r &= \Sigma_{r,N(r)} \Sigma_{N(r)}^{-1} \\
    D_r &= \Sigma_{r,r}  - \Sigma_{r,N(r)} \Sigma_{N(r)}^{-1} \Sigma_{N(i),i}.
\end{split}
\end{equation*}
$\Sigma \in \mathbb{R}^{R\times R}$ is the covariance matrix of $b$ and $\Sigma_{r,N(r)} \in \mathbb{R}^{1\times \tilde{m}}$ denotes the row vector that contains the entries of the $r$-th row of $\Sigma$ that have a column index in $N(r)$, and $\Sigma_{N(r)} \in \mathbb{R}^{\tilde{m}\times \tilde{m}}$ denotes the matrix that contains the entries of $\Sigma$ that have a row and column index in $N(r)$.
Defining the lower triangular matrix $B\in \mathbb{R}^{R\times R}$ with $1$'s on the diagonal, off-diagonal entries
\begin{equation*}
    B_{r,N_{(r)}} = -A_r,
\end{equation*}
and $0$ otherwise, and the diagonal matrix $D\in \mathbb{R}^{R\times R}$ with $D_i$ on the diagonal, we obtain an approximate distribution for the latent Gaussian variable:
\begin{equation*}
    b \overset{\text{approx}}{\sim} \mathcal{N}(0,\tilde\Sigma), ~~~\tilde\Sigma=B^{-1}DB^{-T}.
\end{equation*}
The corresponding precision matrix is sparse and given by $\tilde\Sigma^{-1} = B^TD^{-1}B$.
Calculating the Cholesky factor of $\Sigma_{N(r)}$ is the main computational burden of the approximation. Thus, a Vecchia approximation costs $\mathcal{O}(R\times \tilde{m}^3)$, and saving $B$ requires $O(R \times \tilde{m})$ memory storage. Often, accurate approximations are obtained for small $\tilde{m}$'s.

\clearpage
\subsection{Annual spatial default rates} \label{appendix:default_rates}
\begin{figure}[ht!]
\centering
    \includegraphics[width=0.9\linewidth]{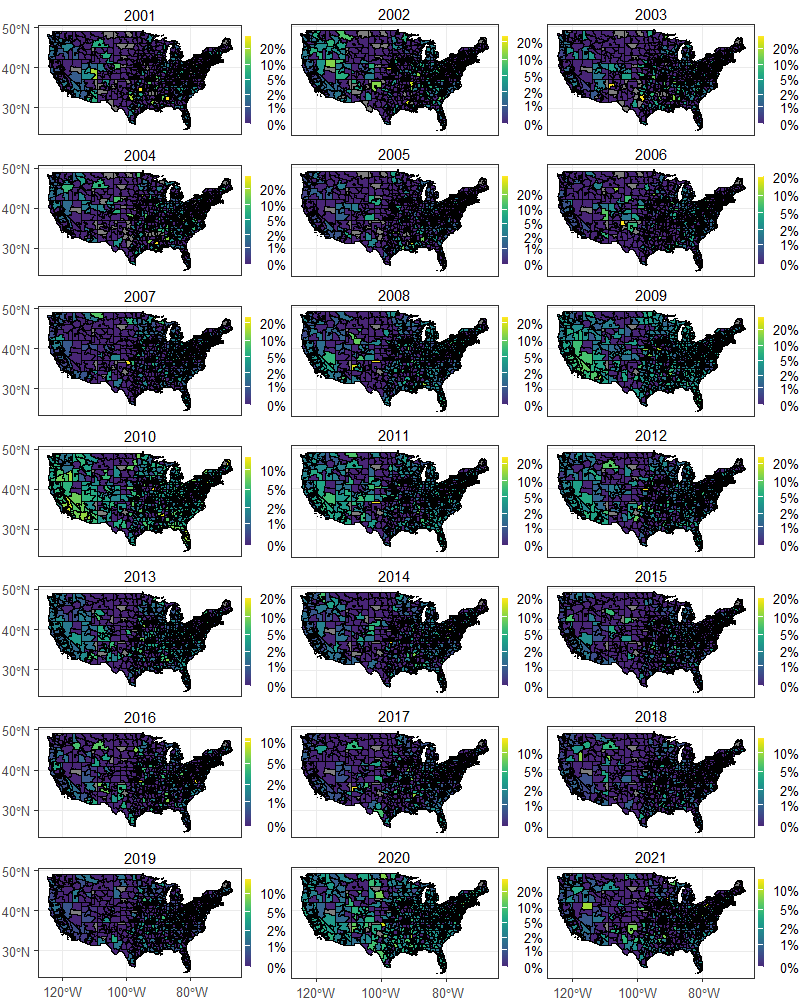}
    \caption{Annual spatial default rates. No default rates are shown for the grey areas because there are too few loan-year observations.}
    \label{fig:yearly_areal_default_rate}
\end{figure}

\clearpage
\subsection{Summary statistics of predictor variables} \label{appendix:summary}
\begin{table}[ht]
\centering
\begin{tabular}{lrrrrrr}
    \hline
        & Min. & Q.25\% & Median & Mean & Q.75\% & Max. \\ 
    \hline
      credit\_score & 300.000 & 698.000 & 743.000 & 734.625 & 779.000 & 850.000 \\ 
      longitude & -123.758 & -104.952 & -87.513 & -92.403 & -80.247 & -67.866 \\ 
      latitude & 25.531 & 35.211 & 39.520 & 38.657 & 41.968 & 48.428 \\ 
      insurance\_percent & 0.000 & 0.000 & 0.000 & 5.344 & 0.000 & 55.000 \\ 
      orig\_dept\_to\_income & 1.000 & 27.000 & 34.679 & 34.339 & 41.000 & 65.000 \\ 
      orig\_loan\_to\_value & 2.000 & 66.000 & 80.000 & 75.735 & 88.000 & 534.000 \\ 
      orig\_unpaid\_balance & 9000.000 & 106000.000 & 160000.000 & 186612.656 & 243000.000 & 1581000.000 \\ 
      curr\_loan\_to\_value & 0.000 & 59.252 & 72.373 & 69.203 & 79.351 & 526.587 \\ 
      ir\_spread & -3.140 & -0.010 & 0.610 & 0.693 & 1.265 & 7.080 \\ 
      n\_months & 0.000 & 12.000 & 29.000 & 39.967 & 57.000 & 274.000 \\ 
      gdp\_growth & -15.019 & 2.539 & 3.994 & 3.951 & 5.578 & 25.184 \\ 
      income\_per\_capita & 9.937 & 10.522 & 10.693 & 10.699 & 10.867 & 11.490 \\ 
      expend\_per\_capita & 9.723 & 10.332 & 10.484 & 10.477 & 10.621 & 11.282 \\ 
      unemployment\_rate & 2.100 & 4.400 & 5.400 & 6.021 & 7.300 & 13.500 \\ 
      hpi\_growth & -23.070 & -0.600 & 3.460 & 3.099 & 6.450 & 31.710 \\ 
      construction\_growth & -33.149 & -0.570 & 5.345 & 3.724 & 8.895 & 42.042 \\ 
      op\_surplus\_growth & -22.903 & 2.150 & 4.630 & 4.561 & 6.569 & 34.548 \\ 
      inflation\_rate & -0.356 & 1.465 & 2.069 & 2.147 & 3.157 & 4.698 \\ 
      djia\_growth & -33.840 & -0.610 & 7.260 & 7.449 & 18.820 & 26.500 \\ 
    \hline
\end{tabular}
\caption{Summary statistics for the numeric predictor variables.} 
\label{table:summary_numeric}
\end{table}
\begin{table}[ht]
    \centering
    \begin{tabular}{lll}
      \hline
            & Level & Count \\ 
      \hline
        occupancy & I & 167551 \\ 
            & P & 2008823 \\ 
            & S & 80154 \\ 
        nr\_units & 1 & 2181346 \\ 
           & 2 & 55477 \\ 
           & 3 & 10304 \\ 
           & 4 & 9401 \\ 
        loan\_purpose & C & 583483 \\ 
           & N & 763290 \\ 
           & P & 909755 \\ 
        first\_time\_homebuyer & 0 & 1972357 \\ 
           & 1 & 284171 \\ 
        msa & 0 & 395311 \\ 
            & 1 & 1861217 \\ 
        multiple\_borrowers & 0 & 1027007 \\ 
            & 1 & 1229521 \\ 
       \hline
    \end{tabular}
\caption{Summary statistics for the categorical predictor variables.}
\label{table:summary_cat}
\end{table}

\clearpage
\subsection{Tuning parameters}\label{tune_pars}
\begin{table}[ht!]
    \centering
    \begin{tabular}{m{6cm} m{3cm}}
        \hline
        Tuning parameter & Candidate values\\ 
        \hline
        Number of trees & \{1,2,...,1000\} \\ 
        Learning rate & \{10,1,0.1\} \\
        Maximal tree depth & \{2,3,5,10\} \\
        Minimal number of samples per leaf & \{10,100,1000\} \\
        L2 regularization & \{0,1,10\} \\
        \hline
    \end{tabular}
    \caption{Candidate tuning parameters for the tree-boosted spatial and spatio-temporal frailty models.} 
    \label{table:candidatesLaGaBoost}
\end{table}
\begin{figure}[ht!]
\centering
   \includegraphics[width=0.9\linewidth]{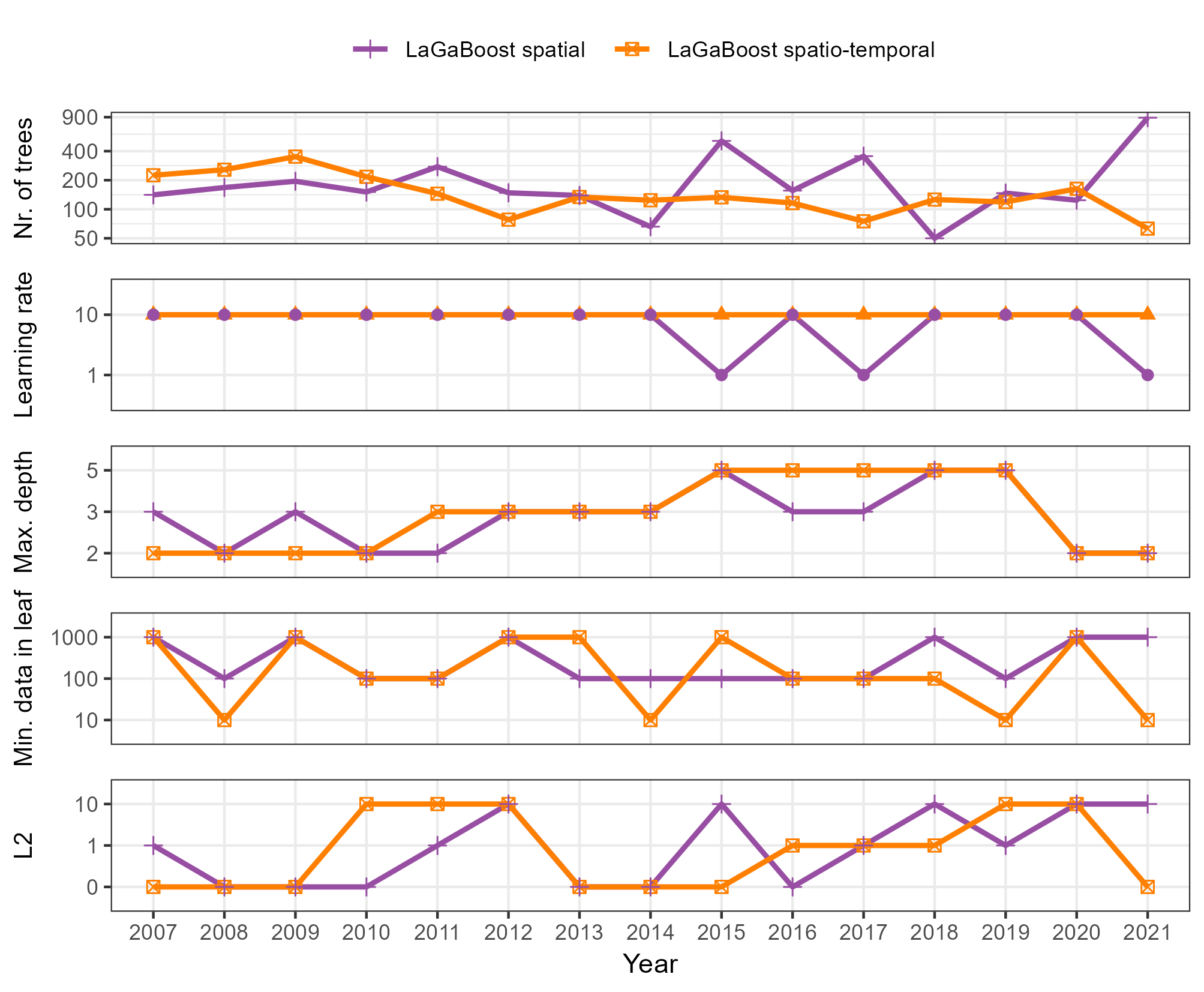}
    \caption{Selected tuning parameters.}
    \label{fig:tuning_pars}
\end{figure}

\clearpage
\subsection{Runtimes} \label{appendix:runtimes}
\begin{table}[ht]
    \centering
    \begin{tabular}{lr}
        \hline
            & Time (s)\\ 
        \hline
            Linear independent & 15.03\\ 
            Linear spatial & 11.54\\ 
            Linear spatio-temporal & 224.48\\ 
            LaGaBoost spatial & 49.34\\ 
            LaGaBoost spatio-temporal & 1637.72\\ 
        \hline
    \end{tabular}
    \caption{Average wall clock times (s) for training on data up to the year 2014.}
    \label{table:runtimes}
\end{table}

\subsection{Additional results for prediction of individual default probabilities}\label{add_res_ind}
\begin{figure}[ht!]
\centering
    \includegraphics[width=0.9\linewidth]{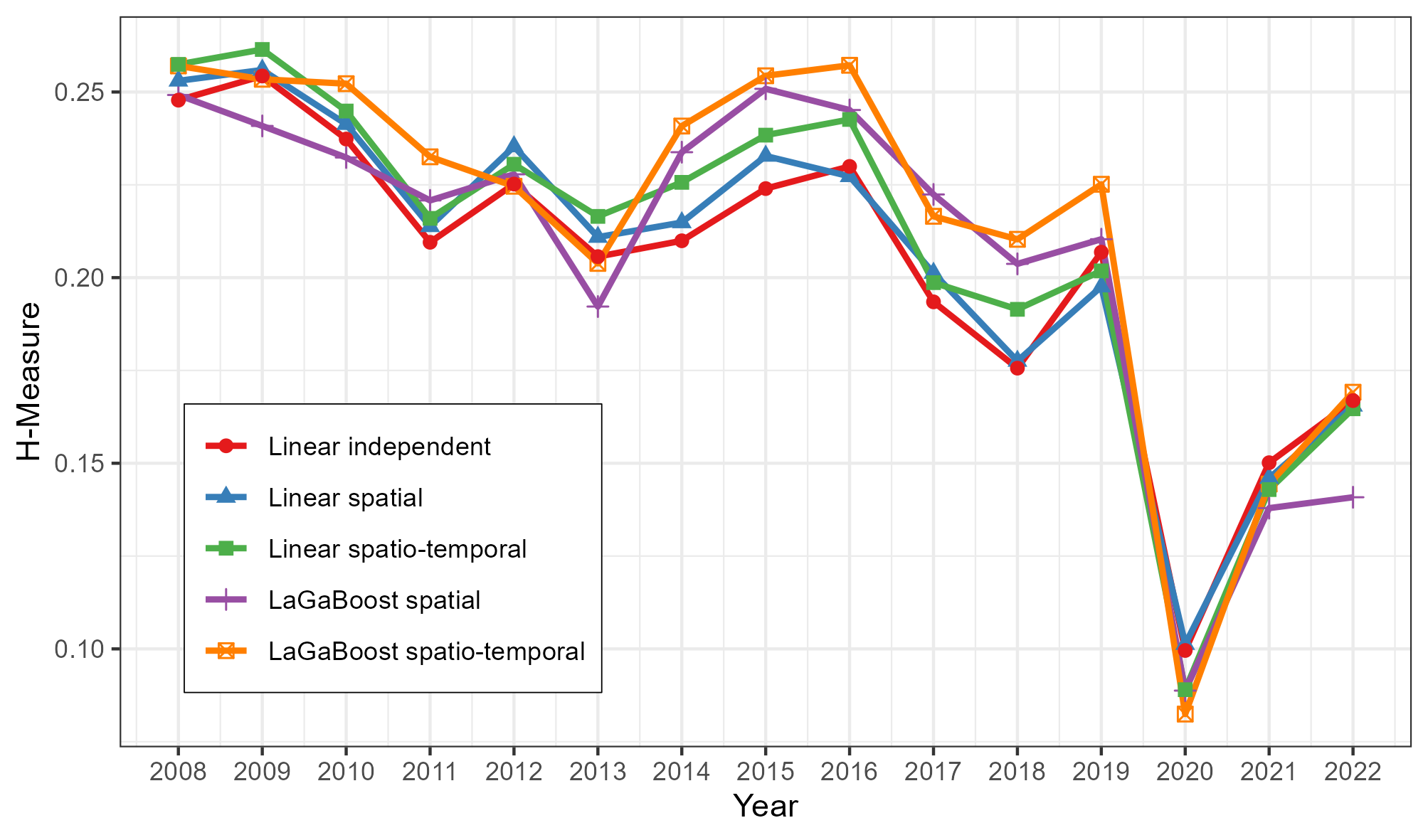}
    \caption{Temporal out-of-sample test H-measure (higher = better).}
    \label{fig:H_measure}
\end{figure}
\begin{figure}[ht!]
\centering
    \includegraphics[width=0.9\linewidth]{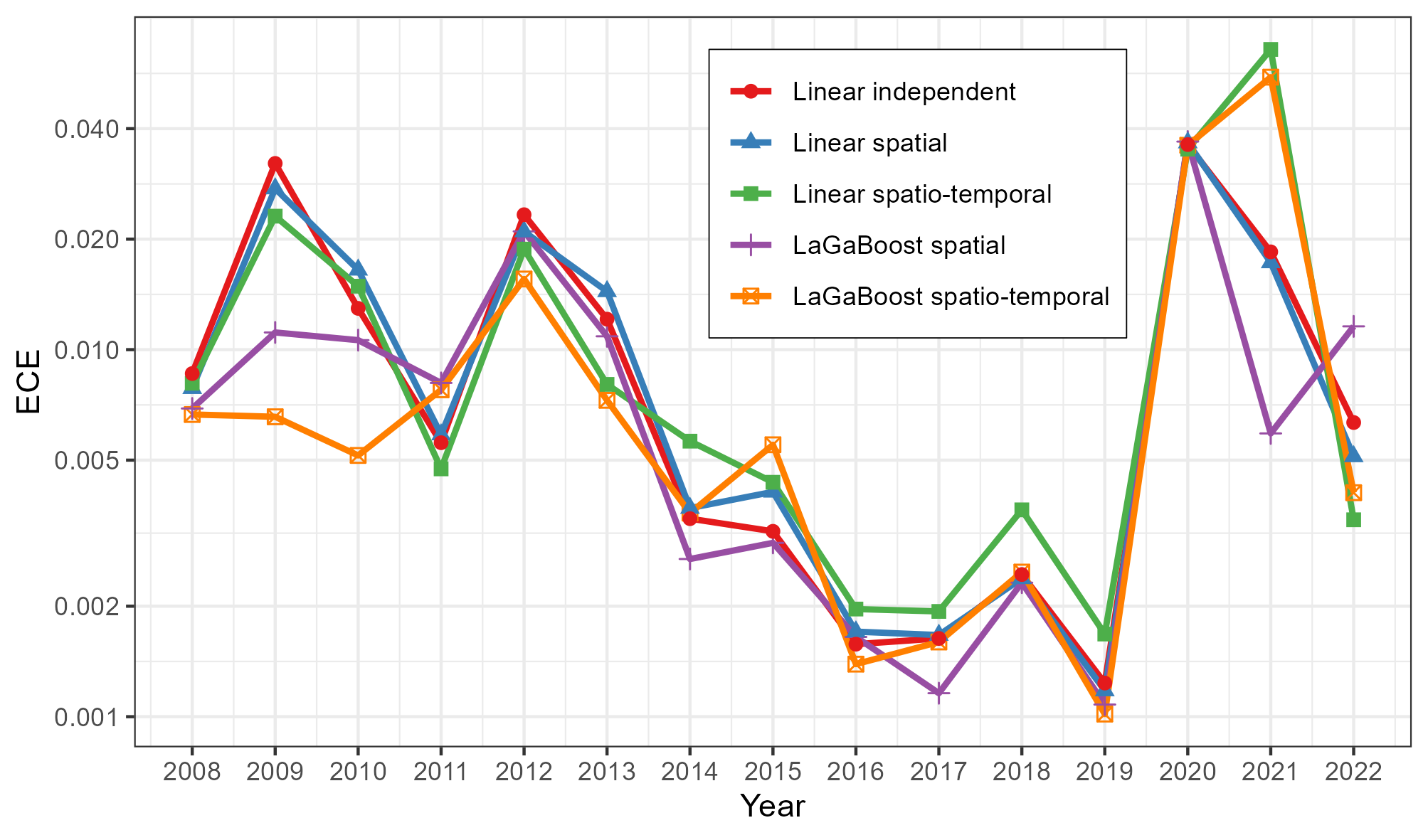}
    \caption{Temporal out-of-sample test Expected Calibration Error (ECE) (lower = better).}
    \label{fig:ECE}
\end{figure}
\begin{table}[ht]
\centering
\begin{tabular}{lr}
  \hline
    Compared models & P-value \\ 
  \hline
        LaGaBoost spatio-temporal vs. LaGaBoost spatial & 1.27e-02 \\
        LaGaBoost spatio-temporal vs. Linear spatio-temporal & 7.56e-27 \\ 
        LaGaBoost spatio-temporal vs. Linear spatial & 1.83e-07 \\ 
        LaGaBoost spatio-temporal vs. Linear independent & 6.76e-09 \\
        LaGaBoost spatial vs. Linear spatio-temporal & 7.78e-09 \\ 
        LaGaBoost spatial vs. Linear spatial & 2.94e-03 \\ 
        LaGaBoost spatial vs. Linear independent & 1.24e-03 \\ 
        Linear spatio-temporal vs. Linear spatial & 2.24e-09 \\ 
        Linear spatio-temporal vs. Linear independent & 1.63e-08 \\ 
        Linear spatial vs. Linear independent & 3.05e-01 \\ 
   \hline
\end{tabular}
    \caption{P-values of the pairwise DeLong tests using the function \textit{roc.test} in the R package \textit{pROC}. AUCs are based on the pooled predictive probabilities of each model over all prediction years.}
    \label{table:DeLong}
\end{table}

\clearpage
\subsection{Additional results for prediction of loan portfolio loss distributions}\label{appendix:loss_distribution}
\begin{figure}[ht!]
\centering
    \includegraphics[width=0.9\linewidth]{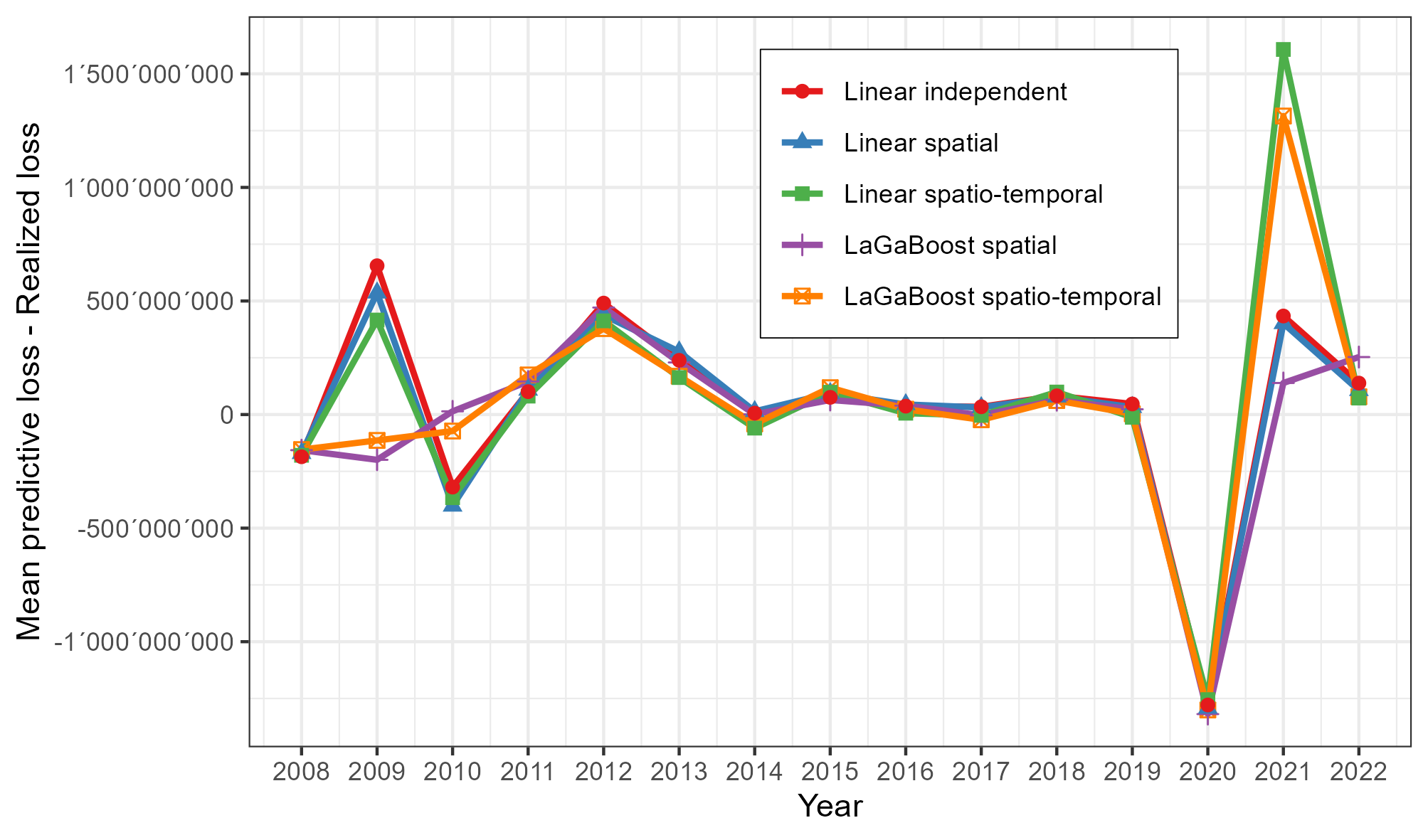}
    \caption{Differences between means of the predictive loss distributions and the realized portfolio losses.}
    \label{fig:diff_loss_all}
\end{figure}

\clearpage
\subsection{Posterior mean for the latent Gaussian process of the linear spatio-temporal model}\label{post_mean_lin}
\begin{figure}[ht!]
\centering
   \includegraphics[width=\linewidth]{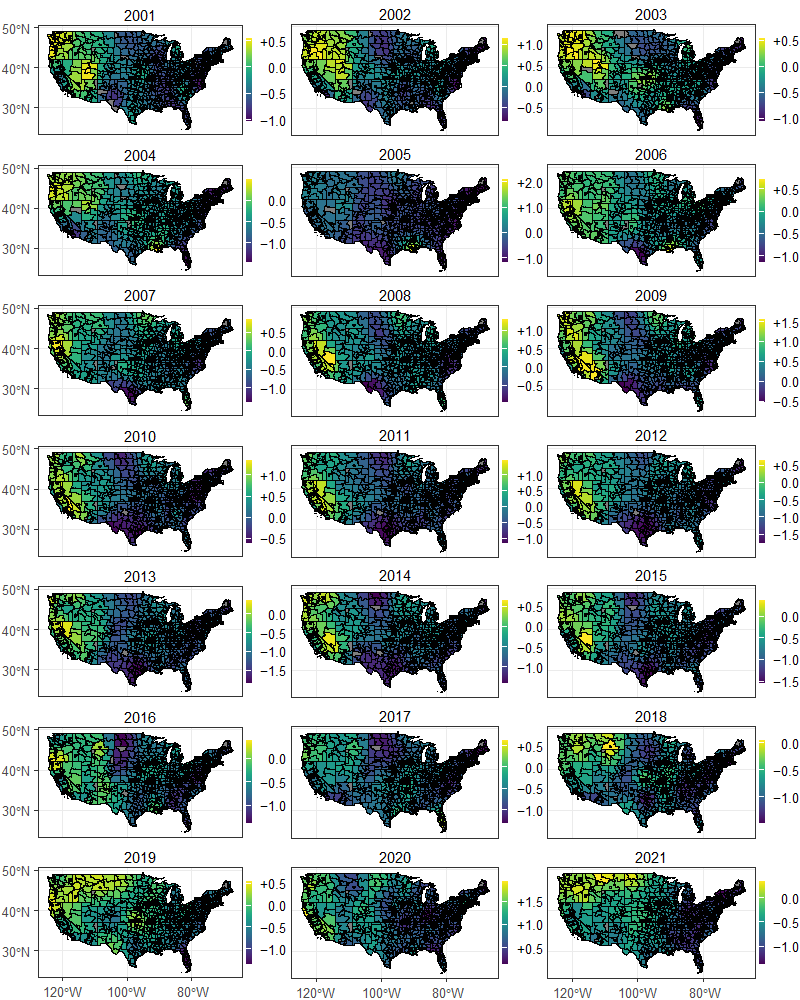}
    \caption{Posterior mean for the latent Gaussian process in the linear spatio-temporal model when training on data up to the year 2021.}
    \label{fig:latent_b_linear}
\end{figure}
\end{document}